\documentclass[11pt,british]{article}
\usepackage[utf8]{inputenc}
\usepackage{jheppub}

\usepackage{mathtools}
\usepackage{amsmath}
\usepackage{amssymb}
\usepackage{cancel}
\usepackage{stmaryrd}
\usepackage{graphicx}
\usepackage{setspace}
\usepackage{microtype}
\usepackage{color}
\usepackage{mathrsfs}
\usepackage{tikz-cd}    							% commutative diagrams
\usepackage{stmaryrd}								% extra symbol font (dbl bracket)
\usepackage[linecolor=blue,backgroundcolor=blue!25,bordercolor=blue]{todonotes}
\presetkeys{todonotes}{inline}{}
	\SetSymbolFont{stmry}{bold}{U}{stmry}{m}{n}		% removes error message for bold dbl bracket font
\SetTracking{encoding={*}, shape=sc}{0} 	% Correct spacing of smallcaps

\usepackage{setspace}
\allowdisplaybreaks		% Allow pagebreaks in multiline eqs

\date{\today} 		% Show today's date

\makeatletter
\gdef\@fpheader{\ }                    % hack the jhep header
\makeatother

\usepackage{slashed}
\usepackage{amsthm}
\usepackage{bm}% bold math

\makeatletter
\g@addto@macro\bfseries{\boldmath}
\makeatother

\def\bar#1{\overline{#1}}

\def\inv{^{\raise.15ex\hbox{${\scriptscriptstyle -}$}\kern-.05em 1}}
\def\lbar{{\lower.35ex\hbox{$\mathchar'26$}\mkern-10mu\lambda}} %lambda bar

\let\p=\partial
\let\<=\langle
\let\>=\rangle

\let\+=\uparrow

\def\a{\alpha'}

\def\bp{\bar{\partial}}
\def\OO{\mathcal{O}}

\def\Q{\mathcal{Q}}

\DeclareMathOperator{\vol}{\text{vol}}
\DeclareMathOperator{\tr}{\text{tr}}
\DeclareMathOperator{\End}{\text{End}}

\DeclareMathOperator{\image}{\text{im}}

\newcommand{\eqspace}{\mathrel{\phantom{=}}{}} % correct spacing after equals for long equations
\newcommand{\ext}{\mbox{\large $\wedge$}} % Exterior algebra
\newcommand{\dd}{\text{d}} % Exterior derivative
\def\d{\text{d}} % Exterior derivative
\newcommand{\del}{\partial} % Exterior derivative
\newcommand{\ee}{\text{e}} % Exponential
\newcommand{\ii}{\text{i}} % Imaginary number
\newcommand{\rep}[1]{\boldsymbol{#1}} % Group representations
\newcommand{\repp}[2]{(\rep{#1},\rep{#2})} % Group representations
\newcommand{\bbZ}{\mathbb{Z}} % Integers
\newcommand{\bbR}{\mathbb{R}} % Reals
\newcommand{\bbC}{\mathbb{C}} % Complex

\newcommand{\SU}[1]{\text{SU}(#1)}
\newcommand{\SO}[1]{\text{SO}(#1)}
\newcommand{\GL}[1]{\text{GL}(#1)}
\newcommand{\HH}{\text{H}}

\theoremstyle{definition}

\newcommand{\ra}{\rightarrow}

\newcommand{\Ra}{\Rightarrow}

\newcommand{\cC}{\mathcal{C}}
\newcommand{\cU}{\mathcal{U}}
\newcommand{\dch}{\partial}
\newcommand{\hs}[1]{\hspace{#1}}

\newcommand{\gtf}{\chi}
\newcommand{\gtft}{\tilde\chi}
\newcommand{\tm}{\tilde{m}}
\newcommand{\hE}{\hat{E}}
\newcommand{\ba}{\bar{a}}

\begin{document}

\title{Finite deformations from a heterotic superpotential: holomorphic Chern--Simons and an $L_\infty$ algebra}% Force line breaks with \\

\author[a]{Anthony Ashmore,}
\author[a]{Xenia de la Ossa,}
\author[b]{Ruben Minasian,}
\author[c,d]{Charles Strickland-Constable}
\author[e,f]{and Eirik Eik Svanes}

\affiliation[a]{Mathematical Institute, University of Oxford, Andrew Wiles Building, \\Radcliffe Observatory Quarter, Woodstock Road, Oxford, OX2 6GG, UK}
\affiliation[b]{Institut de Physique Th\'eorique, Universit\'e Paris Saclay, CEA, CNRS, F-91191, \\Gif-sur-Yvette, France}
\affiliation[c]{School of Mathematics, University of Edinburgh, James Clerk Maxwell Building,\\ Peter Guthrie Tait Road, Edinburgh, EH9 3FD, UK}
\affiliation[d]{School of Physics, Astronomy and Mathematics, University of Hertfordshire, \\College Lane, Hatfield, AL10 9AB, UK}
\affiliation[e]{Department of Physics, King's College London, London, WC2R 2LS, UK}
\affiliation[f]{The Abdus Salam International Centre for Theoretical Physics, 34151 Trieste, Italy}

\subheader{\hfill\textrm{EMPG-18-12}\\ \phantom{A}\hfill\textrm{IPhT-t18/088}}

\null\vskip10pt

\date{\today}% It is always \today, today,

\abstract{We consider finite deformations of the Hull--Strominger system. Starting from the heterotic superpotential, we identify complex coordinates on the off-shell parameter space. Expanding the superpotential around a supersymmetric vacuum leads to a third-order Maurer--Cartan equation that controls the moduli. The resulting complex effective action generalises that of both Kodaira--Spencer and holomorphic Chern--Simons theory. The supersymmetric locus of this action is described by an $L_3$ algebra.}

\keywords{}%Use showkeys class option if keyword
\maketitle

\section{Introduction}

A full understanding of the parameter space of string theory is an outstanding mathematical challenge and would lead to powerful constraints on the landscape of string models. Of the various limits of string theory, the heterotic string has been the focus of much phenomenology thanks to the relative ease with which one can engineer four-dimensional theories with chiral fermions and the Standard Model gauge group~\cite{BCD+12,BHO+05,BHO+05b,BHOP06,AGL+11,AGL+11b,Anderson:2012yf,Anderson:2011cza}. Much of this work has been on models where the internal manifold is Calabi--Yau, mostly because such spaces can be constructed using algebraic geometry and then used for compactifications without knowledge of their explicit metrics. 

Calabi--Yau compactifications are not the most general way to obtain an $\mathcal{N} = 1$ theory in four dimensions that admits a Minkowski vacuum. The general solution to $\mathcal{O}(\alpha')$ is given by compactifying on a complex three-fold $X$ with $H$ flux and a gauge bundle that satisfy an anomaly cancellation condition. The conditions on the geometry and fluxes for such a solution are known as the Hull--Strominger system~\cite{Hull86,Strominger86}. Known solutions to this system include Calabi--Yau spaces with bundles and a small number of honestly non-K\"ahler geometries. Generically, a given solution of the Hull--Strominger system will admit deformations of the geometry, flux and bundle that remain $\mathcal{N}=1$ solutions -- these deformations are known as moduli. These moduli appear in the massless spectrum of the low-energy theory, so it is important that we understand the moduli space of a given compactification.

The moduli spaces of Calabi--Yau compactifications at zeroth order in $\alpha'$ are well understood using the language of special geometry. Until recently the general case had not been tackled -- this might come as a surprise. Certainly in type II theories the conditions for an $\mathcal{N}=1$ Minkowski solution are sufficiently complicated (thanks to branes and other ingredients) that their moduli spaces might not admit a general formulation. In the heterotic case, the underlying geometry is relatively straightforward. One might have expected that the gauge sector and anomaly conditions complicate matters somewhat, but that the moduli space might still be understood. Starting with \cite{AGL+09, AGL+11,Anderson:2011ty,MS11,OS14b,AGS14}, this gap is now being filled. (See also \cite{KMM+11,BMP14b,Bertolini16,BP17} for worldsheet approaches.)

Infinitesimally, the moduli space is characterised by the existence of a holomorphic structure $\bar{D}$ on a bundle $\mathcal{Q}$ over the three-fold $X$. The fact that the moduli space is finite dimensional is intimately connected to this holomorphic structure and the Bianchi identity for the flux. The infinitesimal moduli are captured by the cohomology $\HH_{\bar{D}}^{(0,1)}(\mathcal{Q})$, where $\mathcal{Q}$ is defined by a series of extensions. In this way, the complex structure, hermitian and bundle moduli are combined in a single structure. Furthermore, one can define the analogue of special geometry for these general heterotic compactifications and find the metric on the moduli space~\cite{COM17,McOrist16}.

A natural question to ask is whether one can understand the moduli spaces to higher order. If we think about deformations of a complex structure, we know the infinitesimal moduli are given by $\text{H}^{(0,1)}_{\bar\partial}(T^{(1,0)}X)$, while the higher-order deformations satisfy a Maurer--Cartan equation. A similar thing happens for bundle deformations~\cite{Witten86}, and simultaneously deformations of the bundle and complex structure~\cite{Atiyah57,Huang95}. In this way, moduli can be obstructed at higher orders and can give non-zero contributions to the superpotential of the four-dimensional effective theory. The aim of this work is to derive the corresponding conditions on the moduli for the Hull--Strominger system at higher orders. In other words, we want to derive the conditions on the moduli when they describe a small but \emph{finite} deformation of the original heterotic solution.

There are a number of ways one might go about this. One path would be to start with the equations of the Hull--Strominger system and deform the various fields. The deformed fields should still satisfy the Hull--Strominger system (as it describes the most general solution) so one can rewrite the system of equations as conditions on the deformations themselves. This is similar to the path taken in \cite{COM17,McOrist16}. Our approach will be complimentary. It has been shown that supersymmetry of the heterotic system can be described using a four-dimensional superpotential~\cite{GLM04,OHS16,McOrist16}. The vanishing of the superpotential and its first derivative imposes the $F$-term conditions in the four-dimensional theory and leads to an $\mathcal{N}=1$ Minkowski vacuum. Our plan is to deform the fields that appear in the superpotential and then read off the conditions on the moduli for the superpotential and its derivative to vanish. These two approaches will be shown to be equivalent in a future publication~\cite{AOM+}. A particular advantage of proceeding this way is that one can use the knowledge that the superpotential is a holomorphic function of the moduli fields to streamline the problem. 

In addition to the usual $\mathcal{N}=1$ lore that the superpotential is holomorphic, we give an argument that the superpotential is holomorphic on the space of moduli fields without requiring that they give a solution to the Hull--Strominger system. This is equivalent to saying that the off-shell parameter space -- the space of $\SU{3}$ structures, $B$ fields and gauge bundles -- is a complex space and the superpotential is a holomorphic function of these parameters. We outline how this follows from generalised geometry where $\mathcal{N}=1$ NS-NS compactifications are described by a generalised $\SU{3}\times\SU{4}$ structure~\cite{CSW14b}. We show that the invariant object which characterises this structure is holomorphic and that the complex coordinates on the space of structures match with the usual complex structure, complexified hermitian and bundle deformations.

We show that the conditions on the moduli fields from the vanishing of the superpotential and its first derivative can be written as a pair of third-order Maurer--Cartan equations using the holomorphic structure and a number of brackets. Moreover, we show that the superpotential itself can be rewritten using these operators in a Chern--Simons-like form:
\begin{equation}
\Delta W=\int_{X}\langle y,\bar{D}y-\tfrac{1}{3}[y,y]-\partial b\rangle\wedge\Omega,
\end{equation}
where $y$ describes the complex, hermitian and bundle moduli, and $b$ is a $(0,2)$-form. $\Delta W$ is written using a holomorphic structure $\bar{D}$, a pairing $\langle\cdot,\cdot\rangle$ on the moduli fields, and a bracket $[\cdot,\cdot]$. The brackets can be understood as coming from an underlying holomorphic Courant algebroid that describes the combined deformations of the complex structure, metric and fluxes~\cite{Gualtieri10b,Garcia-Fernandez:2018emx}.

We then show that the supersymmetry conditions can be recast in terms of an $L_3$ algebra~\cite{Zwiebach:1992ie}. We outline how the $L_3$ algebra gives a $\mathcal{C}^\infty$ resolution of the underlying holomorphic Courant algebroid. The natural $L_3$ field equation reproduces the supersymmetry conditions, and the $L_3$ structure gives the gauge symmetries of the moduli space in a compact form.

It is known that generic deformation problems have a description in terms of $L_\infty$ structures, so it is not unexpected that our moduli fields are governed by one. What is unexpected is that the structure truncates at finite order leaving us with an $L_3$ algebra. Why does the deformation truncate in our case? A generic deformation problem can be parametrised in many equivalent ways -- some may truncate at finite order while others do not. Essentially, the structure of the heterotic system and its formulation using a superpotential guides us to pick a ``nice'' parametrisation. Said another way, we know from supergravity that the superpotential should be a holomorphic function of the moduli. Thus when we express the superpotential in the obvious complex coordinates on moduli space, we get the most natural way to package the deformation problem.

We begin in section \ref{sec:HS} with a review of the Hull--Strominger system and the description of its infinitesimal moduli in terms of a holomorphic structure as in \cite{OS14b}. In section \ref{sec:par_defs} we discuss the off-shell parameter space of the theory and give the complex coordinates on the moduli space. We show how the $F$-term conditions follow from a heterotic superpotential to set the scene for the higher-order deformations. In section \ref{sec:HigherDefs} we carry out the higher-order deformation of the superpotential and find the system of equations that govern the moduli of the Hull--Strominger system. We show how this can be written in terms of the holomorphic structure $\bar D$ and a bracket $[\cdot,\cdot]$ arising from a holomorphic Courant algebroid. In section \ref{sec:L3} we rewrite the equations that govern the moduli in terms of an $L_3$ structure. We give the various multilinear products $\ell_k$ that define the $L_3$ structure and discuss how various properties, such as the moduli equations and gauge symmetries, are naturally encoded in this $L_3$ language. In section \ref{sec:ReducedL3} we discuss how the system simplifies under various assumptions and comment on how the effective field theory is encoded in our language. We finish with a discussion of some open questions and avenues for future work.

In the appendices, we lay out our conventions, include a few comments on how flux quantisation works in the heterotic theory, discuss the off-shell parameter space in terms of generalised geometry, show that the $D$-term conditions do not affect the moduli problem and review how the massless moduli are captured by the the cohomology of the holomorphic structure.

%%%%%%%%%%%%%%%%%%%%%%%%%%%%%%%%%%%%%%%%%%%%%%%%%%%%%%%%%%%%%%%%%%%%%%%%%%

\section{The Hull--Strominger system and a heterotic superpotential}\label{sec:HS}

We begin with a review of the Hull--Strominger system~\cite{Hull86,Strominger86} and the description of its infinitesimal moduli using a holomorphic structure~\cite{OS14b,AGS14,GRT15}. 

\subsection{\texorpdfstring{$\mathcal{N}=1$}{N = 1} heterotic vacua and the Hull--Strominger system}

The Hull--Strominger system is a set of equations whose solutions are supersymmetric Minkowski vacua of heterotic string theory to order $\mathcal{O}(\alpha')$. The ten-dimensional solution is a product of four-dimensional Minkowski space with a six-dimensional complex manifold $X$. $X$ admits a vector bundle $V$ with connection $A$ whose curvature $F$ is valued in $\End V$. The tangent bundle $TX$ of $X$ also admits a connection $\Theta$ whose curvature $R$ is valued in $\End TX$. $X$ admits an $\SU3$ structure defined by a nowhere vanishing spinor $\eta$ or, equivalently, a non-degenerate two-form $\omega$ and a nowhere vanishing three-form $\Psi$ that are compatible
\begin{equation}\label{eq:su3}
\omega \wedge \Psi = 0, \qquad \frac{\ii}{\Vert \Psi \Vert^2} \Psi \wedge \bar \Psi = \frac{1}{3!} \omega \wedge \omega \wedge \omega.
\end{equation}
The invariant objects are defined by bilinears of the spinor as
\begin{equation}
\omega_{mn} = -\ii\,\eta^\dagger \gamma_{mn} \eta,\qquad \Psi_{mnp} = \eta^\text{T}\gamma_{mnp}\eta,
\end{equation}
where we are free to normalise the spinor so that $\Vert\Psi\Vert^2=8$. In what follows it will be useful to define a three-form $\Omega$ which is related to $\Psi$ by a dilaton factor as
\begin{equation}
\Omega = \ee^{-2\phi} \Psi.
\end{equation}

Supersymmetry of the vacuum follows from the vanishing of the supersymmetry variations of the fermionic fields, given in equations \eqref{eq:SUSY1}--\eqref{eq:SUSY3}. To first order in $\alpha'$, these conditions are equivalent to the Hull--Strominger system:
\begin{align}
\dd \Omega &= 0, \label{eq:HS1}\\
\ii (\partial - \bar \partial) \omega& = H \coloneqq \dd B + \frac{\alpha'}{4}(\omega_\text{CS}(A)-\omega_\text{CS}(\Theta)), \label{eq:HS2}\\
\Omega \wedge F &= 0, \label{eq:HS3}\\
\omega \lrcorner F &=  0, \label{eq:HS4}\\
\dd (\ee^{-2\phi} \omega \wedge \omega) &= 0, \label{eq:HS5}
\end{align}
where $\omega_\text{CS}$ is the Chern--Simons three-form for the connection,
\begin{equation}
\omega_{\text{CS}} (A) = \tr (A \wedge \dd A + \tfrac23 A\wedge A\wedge A).
\end{equation}
The closure of $\Omega$ from \eqref{eq:HS1} implies that the manifold is complex with a holomorphically trivial canonical bundle, while condition \eqref{eq:HS5} tells us that X is conformally balanced. Conditions \eqref{eq:HS3} and \eqref{eq:HS4} mean $V$ is a polystable holomorphic bundle\footnote{More precisely, it is the complex vector bundle $V_\mathbb{C}$ (defined in appendix \ref{sec:hermitian}) that is a holomorphic bundle.} so the curvature satisfies the hermitian Yang--Mills equations. Finally, \eqref{eq:HS2} defines the flux $H$ in terms of the heterotic $B$ field and the anomaly cancellation condition, and links it with the intrinsic torsion of the $\SU3$ structure. The corresponding Bianchi identity is\footnote{The curvature $R$ in the Bianchi identity is the curvature of a connection on $TX$, satisfying its own hermitian Yang--Mills conditions in order for the equations of motion to be fulfilled \cite{Ivanov10}. To $\mathcal{O}(\alpha')$, this connection is $\nabla^-$, given by taking the connection in \eqref{eq:SUSY1} with the opposite sign for $H$.}
\begin{equation}
\dd H = \frac{\alpha'}{4}(\tr F \wedge F - \tr R \wedge R) \label{eq:HS6}.
\end{equation}
This set of conditions defines what one might call a heterotic $\SU3$ structure.

Upon considering the four-dimensional $\mathcal{N} = 1$ theory that would follow from compactifying on such a solution, the Hull--Strominger system naturally splits into $F$- and $D$-term conditions. As discussed in \cite{OHS16}, the $F$-term equations are
\begin{equation}
\begin{split}\label{eq:F_terms}
\dd \Omega &= 0,\\
\ii (\partial - \bar \partial) \omega &= H,\\
\Omega \wedge F &= 0.\\
\end{split}
\end{equation}
It is these equations that the heterotic superpotential reproduces. The remaining equations of the Hull--Strominger system are the conformally balanced condition and the Yang--Mills equations, referred to as the $D$-term equations. 

Modulo certain mild assumptions on the geometry, the infinitesimal deformations are parametrised by the cohomology $\HH^{(0,1)}_{\bar D}(\Q)$, where $\bar D$ and $\Q$ are to be defined below. This cohomology is reviewed in appendix~\ref{app:massless}. Under infinitesimal deformations, the $D$-term equations fix a representative of a certain cohomology class~\cite{OS14b}, and so should be thought of as gauge fixing conditions that do not affect the moduli problem. This is of course expected from the four-dimensional $\mathcal{N}=1$ supergravity point of view~\cite{Weinberg1998,Weinberg:2000cr,FV12}. In appendix \ref{app:Dterms} we show that preserving the $D$-term conditions for finite deformations also amounts to fixing a gauge. One might worry about Fayet--Iliopoulos terms appearing, but these are in fact accounted for by modding out by $\bar D$-exact terms, as shown in \cite{OS14b}. 

\subsection{The Atiyah algebroid and a holomorphic structure}

The vector bundle $V$ is hermitian in agreement with $(0,2)$ supersymmetry on the worldsheet~\cite{Sen86}. The curvature $F$ of the bundle is given by
\begin{equation}
F = \dd A + A \wedge A,
\end{equation}
where $A$ is a one-form connection valued in $\End V$. The exterior derivative on $V$ twisted by $A$ is
\begin{equation}
\dd_{A}\coloneqq \dd + [A,\cdot],
\end{equation}
where the action of the bracket on a $p$-form $\beta$ is
\begin{equation}
[A,\beta] = A \wedge \beta - (-1)^p \beta \wedge A.
\end{equation}
A holomorphic structure on $V$ is fixed by the $(0,1)$ component of $\dd_A$, which we denote $\bar\partial_{A}$. This operator squares to zero if the bundle is holomorphic, that is $F_{(0,2)} = 0$. Moreover, the Bianchi identity for the curvature is simply
\begin{equation}
\bar\partial_{A} F = 0.
\end{equation}

A deformation of the Hull--Strominger system corresponds to simultaneous deformations of the complex structure, hermitian structure and gauge bundle. Taking each of these in isolation is not sufficient. In particular, deformations of the hermitian structure alone lead to an infinite-dimensional moduli space. It is surprising that if one considers the full deformation problem together with the anomaly cancellation condition, one finds a finite-dimensional moduli space. Of course, this is what one would expect from string theory, but the precise way in which this happens is rather remarkable.

As discussed in \cite{OS14b}, the infinitesimal moduli of the Hull--Strominger system are captured by deformations of a holomorphic structure. The holomorphic structure $\bar D$ acts on a bundle $\Q$. Locally $\Q$ is given by
\begin{equation}\label{eq:Q}
\Q \simeq T^{*(1,0)}(X) \oplus \End V \oplus \End TX \oplus T^{(1,0)}X.
\end{equation}
Globally, $\mathcal{Q}$ is defined by an extension\footnote{Full details can be found in \cite{OS14b}.}
\begin{equation}
0 \rightarrow T^{*(1,0)}(X) \rightarrow \mathcal{Q} \rightarrow \mathcal{Q}_1 \rightarrow 0,
\end{equation}
where the bundle $\mathcal{Q}_1$ is defined by
\begin{equation}
0 \rightarrow \End V \oplus \End TX \rightarrow \mathcal{Q}_1 \rightarrow T^{(1,0)}(X) \rightarrow 0.
\end{equation}

The holomorphic structure $\bar D$ on $\Q$ is a derivative\footnote{A similar operator has appeared in the context of generalised K\"ahler geometry \cite{Gualtieri10b}.}
\begin{equation}
\bar{D}\colon \Omega^{(0,p)}(\Q) \rightarrow  \Omega^{(0,p+1)}(\Q),
\end{equation}
where $\bar{D}^2 = 0$ if and only if the Bianchi identities for $H$, $F$ and $R$ are satisfied. The Hull--Strominger system is then equivalent to the data of the extension bundle $\Q$, the nilpotent holomorphic structure $\bar D$, polystability of $V$ and $TX$ and the conformally balanced condition on $X$.

The infinitesimal deformations of the holomorphic structure are simply elements of the $\bar{D}$-cohomology of $\mathcal{Q}$-valued $(0,1)$-forms -- $\HH^{(0,1)}_{\bar{D}}(\mathcal{Q})$. As shown in \cite{OS14b}, this is also the moduli space of heterotic $\SU3$ structures. We give a short review of this in appendix \ref{app:massless}. 

For the rest of the paper, we make a field redefinition to absorb the explicit $\alpha'$ dependence
\begin{equation}\label{eq:alphaprime}
B\rightarrow\frac{\alpha'}{4}B,\qquad\omega\rightarrow\frac{\alpha'}{4}\omega.
\end{equation}
One can restore the proper factors of $\alpha'$ by the inverse transformations. We will also suppress the connection on $TX$ -- we can reintroduce it in what follows by treating $TX$ as part of the gauge bundle and defining the bundle metric on the $TX$ subspace to be negative definite so that the Bianchi identity comes with a negative sign for the $\tr\,R\wedge R$ term.

The main aim of this work is to understand what happens for finite deformations. In particular we will see the holomorphic structure is an important ingredient in describing higher-order deformations. First let us discuss the off-shell parameter space and how the heterotic $\SU3$ structure can be rephrased using a superpotential.

\section{The off-shell parameter space}\label{sec:par_defs}

We now show that a subset of the system corresponding to $F$-term conditions can be derived from a superpotential. We then discuss how deformations of the geometry, flux and bundle can be parametrised using the observation that the superpotential is a holomorphic function.

The four-dimensional effective theory that one finds after compactifying the heterotic string on an $\SU{3}$ structure manifold is controlled by a superpotential~\cite{GLM04,OHS16,McOrist16}. The superpotential $W$ is given in terms of the flux $H$ and the $\SU{3}$ invariant forms by\footnote{Here we have scaled away an overall factor of $\alpha'/4$ that comes from the field redefinition in \eqref{eq:alphaprime}.}
\begin{equation}\label{eq:super_first}
W=\int_{X}(H+\ii\,\dd\omega)\wedge\Omega.
\end{equation}
As we will review, given the $\SU3$ structure relations \eqref{eq:su3} and the definition of $H$ in \eqref{eq:HS2}, $W = \delta W = 0$ reproduces the $F$-term conditions of the Hull--Strominger system~\cite{OHS16}.

Notice that $\delta W = 0$ requires us to vary the superpotential over some space of field configurations. We need to understand what this space is in order to find how the superpotential behaves when we perform a finite deformation of the background fields. We pause briefly to distinguish between this parameter space and the moduli space of solutions to the Hull--Strominger system.

The parameter space or space of field configurations $\mathcal{Z}$ is the space of $\SU3$ structures, $B$ fields and hermitian gauge bundles on the real manifold $X$. The $\SU3$ structure is equivalent to the existence of a nowhere-vanishing spinor so that on this space of field configurations the heterotic theory admits an ``off-shell'' $\mathcal{N}=1$ supersymmetry of the kind discussed in~\cite{GLSW09,GLW06}. These fields do not necessarily solve the Hull--Strominger system and so we often refer to them as off-shell field configurations. This is the space over which the superpotential is varied.

The moduli space $\mathcal{M}$ of the Hull--Strominger system is a subspace of $\mathcal{Z}$ on which the fields also solve the Hull--Strominger system. This set of fields is what one usually means by moduli, and we will often refer to them as on-shell configurations. Another way of saying this is that the superpotential and its derivatives vanish when evaluated on $\mathcal{M}$.

As $W$ is a superpotential for a four-dimensional $\mathcal{N}=1$ theory we expect it to be a holomorphic function. The holomorphicity of $W$ is a powerful tool for understanding deformations of the Hull--Strominger system and we will see later how its presence greatly simplifies the problem. On the supersymmetric locus (on-shell), it is known that the superpotential is a holomorphic function of the moduli fields~\cite{McOrist16} -- this is simply the statement that anti-holomorphic derivatives of the superpotential vanish on imposing the $F$-term conditions. Physics goes further than this and insists that $W$ is a holomorphic function of the off-shell field configurations -- the off-shell field space $\mathcal{Z}$ must admit complex coordinates and $W$ must be a holomorphic function of these coordinates. In other words, the three-form $\Omega$ and the particular combinations of $B+ \ii\, \omega$ and $A$ that appear in \eqref{eq:super_first} must be parametrised by these complex coordinates.

We outline a proof that such complex coordinates exist and that $\Omega$ is holomorphic on the parameter space $\mathcal{Z}$ in appendix \ref{app:N=1} using the formalism of generalised geometry. We also discuss how the hermitian structure on $V$ survives off-shell. For completeness, one should really show that $W$ itself can be expressed as a holomorphic function of the object $\tilde\chi$ that we define in appendix \ref{app:N=1} -- we leave this for a future work.

Note that on-shell, $\Omega$ is also holomorphic as a function of the complex coordinates of $X$. When we talk of $\Omega$ being holomorphic we are instead referring to its dependence on the coordinates of the off-shell parameter space.

Let $t$ and $\bar{t}$ denote holomorphic and anti-holomorphic coordinates on the parameter space $\mathcal{Z}$. The corresponding holomorphic and anti-holomorphic translation operators are
\begin{equation}\label{eq:variations}
\Delta=\sum_{n=1}^{\infty}\frac{1}{n!}t^{n}\mathcal{D}_{t}^{n},\qquad\bar{\Delta}=\sum_{n=1}^{\infty}\frac{1}{n!}\bar{t}{}^{n}\mathcal{D}_{\bar{t}}^{n},
\end{equation}
where $\mathcal{D}$ is a covariant derivative on the parameter space~\cite{COM17}.

As we discuss in appendix \ref{sec:hermitian}, off-shell the gauge bundle admits a real hermitian connection valued in $V$. This decomposes into $(1,0)$- and $(0,1)$-forms, with a corresponding decomposition of the Chern--Simons form. Not all components of the Chern--Simons form appear in $W$ (as it is wedged with $\Omega$); only the $(0,1)$-form components of $A$ contribute. It is this component of the connection that is the complex coordinate on the off-shell parameter space. A \emph{holomorphic} deformation of the connection is then given by a $(0,1)$-form valued in $\End V$, which we denote $\alpha$:
\begin{equation}
\label{eq:bundle_def}
A\mapsto A+\Delta A= A+\alpha.
\end{equation}
We show in appendix \ref{sec:hermitian} that one can write the Chern--Simons form using connections valued in $V$, $V_\mathbb{C}$, $\bar{V}_\mathbb{C}$ or a combination of these. From this it is clear that it is equivalent to work with $(0,1)$-forms valued in $\End V$ or with $(0,1)$-forms valued in $\End V_\mathbb{C}$.

A deformation of the complex structure is parametrised by a $(1,0)$-vector valued $(0,1)$-form, $\mu\in\Omega^{(0,1)}(T^{(1,0)}X)$, also known as a Beltrami differential. The complex coordinates on $X$ that define the complex structure deform to
\begin{equation}
\dd z^{a}\rightarrow \dd z^{a}+\Delta\dd z^{a}=\dd z^{a}+\mu^{a}.\label{eq:beltrami}
\end{equation}
Infinitesimally, the deformed complex structure is $J+2\ii\,\mu$. For a small but finite deformation, the holomorphic three-form becomes~\cite{Todorov89,BCO+94}
\begin{equation}\label{eq:Omega_def}
\Omega\rightarrow\Omega+\Delta\Omega=\Omega+\imath_{\mu}\Omega+\tfrac{1}{2}\imath_{\mu}\imath_{\mu}\Omega+\tfrac{1}{3!}\imath_{\mu}\imath_{\mu}\imath_{\mu}\Omega,
\end{equation}
where the variations of $\Omega$ in coordinates are
\begin{equation}
\begin{split}\imath_{\mu}\Omega & =\tfrac{1}{2}\Omega_{abc}\mu^{a}\wedge\dd z^{bc},\\
\imath_{\mu}\imath_{\mu}\Omega & =\Omega_{abc}\mu^{a}\wedge\mu^{b}\wedge\dd z^{c},\\
\imath_{\mu}\imath_{\mu}\imath_{\mu}\Omega & =\Omega_{abc}\mu^{a}\wedge\mu^{b}\wedge\mu^{c}.
\end{split}
\end{equation}
The fact that a variation of $\Omega$ is completely captured by $\mu$ without needing $\bar\mu$ is an indication that $\Omega$ is a holomorphic function of the coordinates of the off-shell parameter space. Note that the variation of $\Omega$ can in principle have a $(3,0)$ component. However, the $(3,0)$ part should be interpreted as a K\"ahler transformation and so is not part of the physical moduli. Another way of saying this is that $\Delta$ in \eqref{eq:variations} is built from covariant derivatives on the parameter space~\cite{COM17}. Restricted to variations of the complex structure, the $(3,0)$ component is attributed to a connection in the usual way.

The holomorphic deformations of the hermitian and $B$ field moduli are
\begin{equation}\label{eq:complex}
(\mathcal{B}+ \ii \,\Delta\omega)_{(1,1)} \quad\text{and}\quad \mathcal{B}_{(0,2)},
\end{equation}
where a subscript $(p,q)$ denotes the type with respect to undeformed complex structure. Here $\mathcal{B}$ is a combination of variations of the $B$ field and the exact term in the variation of the Chern--Simons term, given by
\begin{equation}
\mathcal{B} = \Delta B + \tr \Delta A \wedge A,
\end{equation}
up to a $\dd$-closed two-form. The Green--Schwarz mechanism ensures $\mathcal{B}$ is gauge invariant. As we show in appendix \ref{sec:flux_quant}, flux quantisation then implies that $\mathcal{B}$ is a globally defined two-form, so it can indeed be a modulus. As we will see, the $(0,2)$ component of $\Delta\omega$ is actually fixed in terms of the other moduli by the $\SU3$ relations, but it will be convenient to package this with $\mathcal{B}_{(0,2)}$ into $(\mathcal{B}+ \ii \,\Delta\omega)_{(0,2)}$.

\subsection{\texorpdfstring{$F$}{F}-term conditions from the superpotential}

Let us review how one derives the $F$-term conditions from the superpotential. We take $\delta$ to be an infinitesimal deformation of the fields, leading to a corresponding variation of the superpotential $\delta W$
\begin{equation}
\delta W=\int_{X}\bigl(2 \tr \delta A \wedge F + \dd (\mathcal{B} + \ii \, \delta \omega)\bigr) \wedge \Omega + \int_X (H + \ii \,\dd \omega) \wedge \delta \Omega.
\end{equation}
The $F$-term conditions come from requiring that both $W$ and $\delta W$ vanish for generic values of the moduli. For arbitrary $\mathcal{B} + \ii \, \delta \omega$ and $\delta A$, the vanishing of $\delta W$ requires
\begin{equation}
\bar{\partial}\Omega=\dd\Omega=0, \qquad F \wedge \Omega = 0.
\end{equation}
With this in mind, the vanishing of $\delta W$ for arbitrary $\delta \Omega$ of type $(2,1)$ implies $H_{(1,2)}=-\ii\,\bar{\partial}\omega$. Using the previous conditions, the vanishing of $W$ itself reduces to $H_{(0,3)}=\bar{\partial}\gamma$ for some $(0,2)$-form $\gamma$. The Bianchi identity for $H$ then implies $H_{(0,3)}=0$, giving us the final $F$-term condition:
\begin{equation}
H=\ii(\partial-\bar{\partial})\omega.
\end{equation}
With this we see $W = \delta W = 0$ reproduces the $F$-term equations of the Hull--Strominger system.

Our plan is to extend this discussion to understand finite deformations around a supersymmetric solution. First we need to understand how the requirement of an $\SU3$ structure and holomorphicity of the superpotential constrain the possible deformations.

\subsection{Constraints from the SU(3) structure and holomorphicity}
\label{sec:constraints-SU3-holomorphicity}

The existence of an $\SU3$ structure is part of the data that goes into the superpotential, so the deformed geometry should also define an \SU3 structure. Another way of saying this is that the off-shell parameter space on which the superpotential is varied is the space of $\SU3$ structures (plus bundles, and so on). This means the $\SU3$ structure compatibility condition must still hold:
\begin{equation}
(\omega+\Delta\omega)\wedge(\Omega+\Delta\Omega)=0,
\end{equation}
where $\Delta$ is a finite holomorphic variation. Expanding this out according to complex type, we find
\begin{equation}
0	=\big(\omega+(\Delta\omega)_{(2,0)}+(\Delta\omega)_{(1,1)}+(\Delta\omega)_{(0,2)}\big)\wedge\big(\Omega+\imath_{\mu}\Omega+\tfrac12\imath_{\mu}\imath_{\mu}\Omega+\tfrac{1}{3!}\imath_{\mu}\imath_{\mu}\imath_{\mu}\Omega\big).
\end{equation}
Upon contracting this with $\Omega$, we see this equation fixes the $(0,2)$ component of $\Delta\omega$ in terms of the other deformations:
\begin{equation}
(\Delta\omega)_{(0,2)}=\imath_{\mu}\omega+\imath_{\mu}(\Delta\omega)_{(1,1)}-\tfrac12\imath_{\mu}\imath_{\mu}(\Delta\omega)_{(2,0)}.
\end{equation}

Now consider an anti-holomorphic variation $\bar{\Delta}$ under which $\Omega$ does not vary as it is a holomorphic on the parameter space. We then note that as an anti-holomorphic deformation does not change the complex structure (as $\bar\Delta \Omega=0$) the $\SU3$ compatibility condition reduces to
\begin{equation}
\bar\Delta\omega\wedge\Omega=0.
\end{equation}
From this we see $(\bar\Delta\omega)_{(0,2)}=0$ and so, taking a conjugate, $(\Delta\omega)_{(2,0)}=0$. Combined with the previous result of a holomorphic variation of the compatibility condition, we have
\begin{equation}
\label{eq:su3_compat}
(\Delta\omega)_{(2,0)}=0,\qquad(\Delta\omega)_{(0,2)}=\imath_{\mu}\omega+\imath_{\mu}(\Delta\omega)_{(1,1)}.
\end{equation}
From this we see the $(0,2)$ component of the variation of $\omega$ is fixed by the complex structure and hermitian moduli. 

We can also play the same trick with the superpotential itself. The superpotential is a holomorphic function of the moduli so an arbitrary anti-holomorphic variation of it must vanish without having to impose the supersymmetry conditions -- it must vanish ``off shell''. Let us turn off the gauge sector for now, and consider an anti-holomorphic variation of $W$:
\begin{equation}\label{eq:super_anti_hol}
\bar{\Delta}W=\int_{X}\dd(\bar{\Delta}B+\ii\,\bar{\Delta}\omega)\wedge\Omega=-\int_{X}(\bar{\Delta}B+\ii\,\bar{\Delta}\omega)\wedge\dd\Omega.
\end{equation}
As $W$ is a holomorphic functional, this must vanish for all anti-holomorphic deformations without needing to impose the $F$-term conditions. Without imposing integrability of the complex structure, generically we have $\dd\Omega\in\Omega^{(3,1)}(X)\oplus\Omega^{(2,2)}(X)$. For an anti-holomorphic variation of the superpotential to vanish, it is sufficient that
\begin{equation}
\label{eq:van_suff}
(\bar{\Delta}B+\ii\,\bar{\Delta}\omega)_{(1,1)}=0,\qquad(\bar{\Delta}B)_{(0,2)}=0,
\end{equation}
where we have used the first condition from \eqref{eq:su3_compat} to remove $(\bar\Delta\omega)_{(0,2)}$. This agrees with \eqref{eq:complex} where we stated that the holomorphic combinations are the $(1,1)$ and $(0,2)$ components of $ B+\ii\,\omega$ (see also discussion in Appendix \ref{app:O66}). Taking a conjugate of these conditions we have
\begin{equation}
(\Delta B-\ii\,\Delta\omega)_{(1,1)}=0,\qquad(\Delta B)_{(2,0)}=0.
\end{equation}
 Taken together these give
\begin{equation}
(\Delta B)_{(1,1)}=\ii\,(\Delta\omega)_{(1,1)},\qquad(\Delta B)_{(2,0)}=(\Delta\omega)_{(2,0)}=0.
\end{equation}
For what follows, it is useful to define 
\begin{equation}\label{eq:bx_def}
\tilde b=(\Delta B+\ii\,\Delta\omega)_{(0,2)},\qquad x=\ii\,(\Delta\omega)_{(1,1)}=(\Delta B)_{(1,1)},
\end{equation}
which are our complex coordinates on the parameter space.

One can repeat this exercise with gauge sector turned on. $W$ is a holomorphic function of the moduli so it does not change for arbitrary anti-holomorphic variations $\bar\Delta A$. For $\bar{\Delta}W=0$ to hold at a generic off-shell point in field space, we find it is sufficient that $(\bar\Delta A)_{(0,1)}=0$. This implies $(\Delta A)_{(1,0)}=0$, so the holomorphic deformations correspond to
\begin{equation}
\Delta A=\alpha\in\Omega^{(0,1)}(\End V).
\end{equation}
In other words, the holomorphic coordinate on the parameter space is $\alpha$, in agreement with \eqref{eq:bundle_def}. Furthermore, one sees that the holomorphic deformations of the complexified hermitian moduli are the (1,1) and (0,2) components of $\mathcal{B} + \ii \Delta \omega$.

As an aside, we note that there is a schematic way to see that $\Omega$ is a holomorphic function of the parameter space coordinates. Consider a generic anti-holomorphic deformation of the superpotential around a point in moduli space where the holomorphic top-form is closed, $\dd\Omega=0$:
\begin{equation}
\label{eq:AntidefW}
\bar\Delta W=\int_X(H+\ii\,\dd\omega)\wedge\bar\Delta\Omega+\int_X\dd(\bar\Delta B+\ii\,\bar\Delta\omega)\wedge\bar\Delta\Omega.
\end{equation}
For an infinitesimal deformation, the second term can be dropped, and a sufficient condition for the first term to vanish for generic $H$ and $\omega$ is that
\begin{equation}
\label{eq:AntidefOm}
\bar\Delta\Omega=0,
\end{equation}
infinitesimally. Note that this also kills the second term in \eqref{eq:AntidefW} at second order in perturbation theory. A sufficient condition for $\bar\Delta W$ to vanish at this order is hence that $\bar\Delta\Omega=0$ to this level as well. This argument can be continued ad infinitum, and we are left with condition \eqref{eq:AntidefOm}, at least for finite deformations away from a closed $\Omega$. We will assume that this condition holds true at generic off-shell points in the parameter space. Stronger evidence for this is provided by the generalised geometry formulation of the off-shell parameter space presented in appendix~\ref{app:N=1}; we also find a matching between the complex coordinates and the natural parametrisation from generalised geometry.

%%%%%%%%%%%%%%%%%%%%%%%%%%%%%%%%%%%%%%%%%%%%%%%%%%%%%%%%%%%%%%%%%%%%%%%%%%

\section{Higher-order deformations}\label{sec:HigherDefs}

The main aim of this paper is to derive the conditions on the moduli when we move from infinitesimal to finite deformations of solutions to the Hull--Strominger system. In other words, we consider higher-order deformations of the fields that parametrise the supersymmetric Minkowski solution. As we have mentioned we only need to consider the $F$-term relations to understand the moduli space. We show in appendix \ref{app:Dterms} that under some reasonable assumptions the $D$-term conditions do not constrain the moduli problem and should be thought of as gauge fixing conditions -- we expect this to hold in general.

\subsection{The superpotential}

Let us consider the effect on the superpotential of a finite deformation of the background fields away from a point on the supersymmetric locus. In other words, we start with a supersymmetric vacuum solution described by a superpotential $W$ which is a functional of the $\SU{3}$ structure, $H$ and the bundle. Let us denote the superpotential evaluated at this point by $W|_0$. The vacuum is supersymmetric if both the superpotential and its first derivative vanish when evaluated on the solution. Now move a finite distance from this solution in parameter space by deforming the background. The superpotential evaluated at this new point is $W|_\Delta = W|_0 + \Delta W$. We have a supersymmetric solution if both $W|_\Delta$ and its first derivative vanish at that point in parameter space, which is equivalent to the vanishing of $\Delta W$ and $\delta \Delta W$. Let us see how this works out. For clarity of presentation, let us ignore the bundle moduli -- we will reinstate these in section \ref{sec:incl-bundle}. 

A finite holomorphic deformation of the parameters gives
\begin{equation}
\begin{split}
\Delta W&=\int_{X}\Bigl[\bigr(H+\ii\,\dd\omega+\dd (\Delta B + \ii \, \Delta \omega)\bigl)\wedge(\imath_{\mu}\Omega+\tfrac12\imath_{\mu}\imath_{\mu}\Omega+\tfrac{1}{3!}\imath_{\mu}\imath_{\mu}\imath_{\mu}\Omega) \\
&\eqspace\phantom{\int_{X}}+\dd (\Delta B + \ii \, \Delta \omega)\wedge\Omega\Bigr].
\end{split}
\end{equation}
As we are deforming about a supersymmetric point we have $H=\ii(\partial-\bar{\partial})\omega$ and $\dd\Omega=0$, so the first term simplifies and the last term vanishes, giving
\begin{equation}
\begin{split}
\Delta W&=\int_{X}\Bigl[\ii\,\partial\omega\wedge\imath_{\mu}\imath_{\mu}\Omega+\dd (\Delta B + \ii \, \Delta \omega)\wedge(\imath_{\mu}\Omega+\tfrac12\imath_{\mu}\imath_{\mu}\Omega)\Bigr]\\
&=\int_{X}\Bigl[\ii\,\partial\omega\wedge\imath_{\mu}\imath_{\mu}\Omega+2\,\bar\partial x \wedge\imath_{\mu}\Omega+\partial x\wedge\imath_{\mu}\imath_{\mu}\Omega+\partial \tilde{b} \wedge\imath_{\mu}\Omega\Bigr],
\end{split}
\end{equation}
where the $\imath_{\mu}\imath_{\mu}\imath_{\mu}\Omega$ term vanishes due to the type of $\dd(\Delta B + \ii \, \Delta \omega)$, and we have written the second line in terms of $\tilde{b}$ and $x$, the $(0,2)$ and $(1,1)$ parts of the complexified hermitian moduli~\eqref{eq:bx_def}. As $\imath_\mu$ satisfies a graded Leibniz identity, we can rewrite the above as
\begin{equation}
\begin{split}\label{eq:action_without_D}
\Delta W&=2\int_{X}(-\imath_{\mu}\bar\partial x+\tfrac12\ii\,\imath_{\mu}\imath_{\mu}\partial\omega+\tfrac12\imath_{\mu}\imath_{\mu}\partial x-\tfrac{1}{2}\imath_{\mu}\partial \tilde{b} )\wedge\Omega \\
&=2\int_{X}(\mu^{d}\wedge\bar{\partial}x_{d}+\ii\,\mu^{d}\wedge\mu^{e}\wedge\partial_{d}\omega_{e\bar{c}}e^{\bar{c}}+\mu^{d}\wedge\mu^{e}\wedge\partial_{d}x_{e}-\tfrac{1}{2}\mu^{d}\wedge\partial_{d}\tilde{b})\wedge\Omega.
\end{split}
\end{equation}
Our first condition for the deformed background to be supersymmetric is $\Delta W = 0$ when evaluated on the solution. In other words, the terms in the brackets in \eqref{eq:action_without_D} should be zero up to a $\bar\partial$-exact term.

We also need to impose the vanishing of the first derivative of $\Delta W$. As $\Delta W$ is a functional, this amounts to treating it as an action and finding the resulting equations of motion. Varying $\Delta W$, one finds
\begin{align}
\bar{\partial}x_{a}-\ii\,\mu^{d}\wedge(\partial\omega)_{da\bar{b}}e^{\bar{b}}+\mu^{d}\wedge\partial_{a}x_{d}-\mu^{d}\wedge\partial_{d}x_{a}-\tfrac{1}{2}\partial_{a}\tilde{b} & =0,\label{eq:x_eom}\\
\bar{\partial}\mu^{d}-\tfrac{1}{2}[\mu,\mu]^{d} & =0,\label{eq:MC}\\ 
\partial\imath_{\mu}\Omega & =0. \label{eq:constraint}
\end{align}

A few comments are in order. The condition in \eqref{eq:MC} is nothing but the Maurer--Cartan equation for finite deformations of a complex structure. This is somewhat expected as we know solutions to the Hull--Strominger system are manifolds with a complex structure. Notice that we also have a second condition on $\mu$ in \eqref{eq:constraint} which is not usually seen in discussions on the moduli space of complex structures. This condition comes from requiring that the deformed three-form $\Omega + \Delta \Omega$ is closed and thus holomorphic with respect to the new complex structure -- this is stronger than requiring a complex structure alone. Note that this same condition that appears in \cite{BCO+94} for Kodaira--Spencer gravity -- there it is imposed as a constraint from the outset but one should actually think of it as requiring that the deformed three-form $\Omega$ remains $\dd$-closed. One could make a change of variables which solves this constraint explicitly by taking $\imath_\mu\Omega = a+\partial b$ where $a$ is $\partial$-harmonic. This may be useful when investigating the quantum theory defined by the superpotential but we do not use it in what follows.

\subsection{A Maurer--Cartan equation from the holomorphic structure}

The idea now is that these equations can be interpreted as a Maurer--Cartan equation for the deformations. We know the infinitesimal moduli of the Hull--Strominger system are captured by the $\bar{D}$-cohomology of the holomorphic structure, so we expect $\bar{D}$ will be the differential that appears in such a Maurer--Cartan equation. The other ingredient is a bracket. We introduce the bundle $\mathcal{Q}'$ as the sum of the holomorphic cotangent and tangent bundles:
\begin{equation}
\mathcal{Q}'\simeq T^{*(1,0)}X \oplus T^{(1,0)}X.
\end{equation}
This is the bundle $\mathcal{Q}$ defined in \eqref{eq:Q} with the gauge sector suppressed for now. We write $(0,1)$-forms valued in this bundle as
\begin{equation}
y\in\Omega^{(0,1)}(\mathcal{Q}'),\qquad y=(x_{a}e^a,\mu^{a}\hat{e}_a) \equiv x + \mu.
\end{equation}
The holomorphic structure for the Hull--Strominger system without the bundles is given by an operator $\bar{D}$ that acts on sections of $\mathcal{Q}'$. We can also introduce a bracket $[\cdot,\cdot]$ on forms valued in $\mathcal{Q}'$ and a pairing $\langle\cdot,\cdot\rangle$ that traces over the $\mathcal{Q}'$ indices:
\begin{align}
[\cdot,\cdot]&\colon \Omega^{(0,p)}(\mathcal{Q}')\times\Omega^{(0,q)}(\mathcal{Q}') \rightarrow \Omega^{(0,p+q)}(\mathcal{Q}'), \\
\langle\cdot,\cdot\rangle&\colon \Omega^{(0,p)}(\mathcal{Q}')\times\Omega^{(0,q)}(\mathcal{Q}') \rightarrow \Omega^{(0,p+q)}(X).
\end{align}
We give explicit expressions for $\bar{D}$, the bracket $[\cdot,\cdot]$ and the pairing $\langle\cdot,\cdot\rangle$ in equations (\ref{eq:Dbar_definition}), (\ref{eq:bracket_definintion}) and (\ref{eq:pairing_definition}). Using these we can rewrite $\Delta W$, given in \eqref{eq:action_without_D}, as
\begin{equation}
\Delta W=\int_{X}\langle y,\bar{D}y-\tfrac{1}{3}[y,y]-\partial b\rangle\wedge\Omega,\label{eq:action_with_b}.
\end{equation}
Note that we have redefined the $(0,2)$-form field as $ b =\tilde{b}-\mu^{a}\wedge x_{a}$. The equations of motion \label{eq:x_eom,eq:MC,eq:constraint} that follow from varying $\Delta W$ can then be written compactly as
\begin{align}
\bar{D}y-\tfrac{1}{2}[y,y]-\tfrac{1}{2}\partial b & =0,\label{eq:y_eom}\\
\partial\imath_{\mu}\Omega & =0.\label{eq:b_eom}
\end{align}

Let us make a few comments. Looking at $\Delta W$ in equation (\ref{eq:action_with_b}), we see it resembles a Chern--Simons action. More specifically, the form of the action is that of a holomorphic Chern--Simons theory for $y$ with a Lagrange multiplier $b$ that enforces a constraint for $y$. This constraint is the same as the gauge choice that is imposed in Kodaira--Spencer theory~\cite{DGNV05,BCO+94}. Note that the conventional Chern--Simons action has appeared as a superpotential in other work~\cite{Witten:2010zr}; we expect a similar analysis could be applied here. 

Notice also that infinitesimal deformations are captured by 
\begin{equation}
\bar{D}y=\tfrac{1}{2}\partial b.
\end{equation}
It follows from this and $\dd H\propto\partial\bar{\partial}\omega=0$ that $\partial b$ is $\bar{\partial}$-closed.\footnote{Recall we have turned off the gauge sector in this subsection.} If the underlying manifold $X$ satisfies the $\partial\bar{\partial}$-lemma or $\text{H}^{(0,2)}(X)$ vanishes, $\partial b$ is $\bar{\partial}$-exact and can be absorbed in a redefinition of the complexified K\"ahler moduli $x$. We then see that infinitesimally the complexified K\"ahler moduli are counted by $\text{H}^{(0,1)}(T^{*(1,0)}X)$.

\subsection{Including the bundle}
\label{sec:incl-bundle}
We now want to include the bundle degrees of freedom in the superpotential -- we do this by adding a Chern--Simons term $\omega_{\text{CS}}(A)$ for the gauge connection $A$:
\begin{equation}
W=\int_{X}\bigl(\dd B+\omega_{\text{CS}}(A)+\ii\,\dd\omega\bigr)\wedge\Omega.
\end{equation}
The $B$ field transforms in the usual Green-Schwarz manner so that $H$ is gauge invariant:
\begin{equation}
\delta_{\epsilon}B=\tr(\dd\epsilon\wedge A).
\end{equation}

Consider a shift of the gauge connection by $A\rightarrow A+\alpha.$ The corresponding change of the Chern--Simons form is
\begin{equation}
\Delta\omega_{\text{CS}}=2\tr(F\wedge\alpha)+\tr(\alpha\wedge\dd_{A}\alpha)+\tfrac{2}{3}\tr(\alpha\wedge\alpha\wedge\alpha)+\dd\tr(\alpha\wedge A).
\end{equation}
The exact term in this variation combines with the variation of the $B$ field to give
\begin{equation}
\mathcal{B}=\Delta B+\tr(\alpha\wedge A).
\end{equation}
As we show in appendix \ref{app:hetGlobal}, $\mathcal{B}$ is a globally defined two-form so it can be a modulus~\cite{OS14b}.

Expanding about a supersymmetric point, the extra terms in the variation of the superpotential from the gauge sector are
\begin{equation}
\begin{split}\Delta W_{\alpha}  & =\int_{X}\Delta\omega_{\text{CS}}\wedge(\Omega+\imath_\mu\Omega+\imath_\mu\imath_\mu\Omega)\\
 & =\int_{X}\Bigl[(\tr(\alpha\wedge\bar{\partial}_{A}\alpha)+\tfrac{2}{3}\tr(\alpha\wedge\alpha\wedge\alpha))\wedge\Omega\\
 &\eqspace\phantom{\frac{\alpha'}{4}\int_{X}}+(2\tr(F\wedge\alpha)+\tr(\alpha\wedge\partial_{A}\alpha))\wedge\imath_\mu\Omega\\
 &\eqspace\phantom{\frac{\alpha'}{4}\int_{X}}+\dd\tr(\alpha\wedge A)\wedge(\imath_\mu\Omega+\tfrac12\imath_\mu \imath_\mu\Omega)\Bigr],
\end{split}
\end{equation}
where the final terms combine with $b$ in $\Delta W$ to give $\mathcal{B}=b+\tr(\alpha\wedge A)$. Upon replacing $b$ by $\mathcal{B}$ in \eqref{eq:action_without_D}, the extra terms in the variation of the superpotential are
\begin{equation}
\Delta W_{\alpha}=\int_{X}\bigl(\tr(\alpha\wedge\bar{\partial}_{A}\alpha)+\tfrac{2}{3}\tr(\alpha\wedge\alpha\wedge\alpha)-2\mu^{d}\wedge F_{d}\wedge\alpha-\alpha\wedge\mu^{d}\wedge(\partial_A\alpha)_d\bigr)\wedge\Omega.\label{eq:action_for_gauge_sector}
\end{equation}
The full expression for $\Delta W$ is given by the sum of expressions (\ref{eq:action_for_gauge_sector}) and (\ref{eq:action_without_D}) (with $b\rightarrow\mathcal{B}$) . The equations of motion that follow from varying the full superpotential are
\begin{equation}
\begin{split}
\bar{\partial}x_{a}-\ii\,\mu^{d}\wedge(\partial\omega)_{da\bar{b}}e^{\bar{b}}-\tr F_a \wedge \alpha+\mu^{d}\wedge\partial_{a}x_{d}-\mu^{d}\wedge\partial_{d}x_{a}+\tfrac{1}{2} \tr \alpha\wedge(\partial_A \alpha)_a-\tfrac{1}{2}\partial_{a}b & =0,\\
\bar{\partial}_{A}\alpha+\alpha\wedge\alpha+F_{d\bar{a}}\dd\bar{z}^{\bar{a}}\wedge\mu^{d}-\mu^{d}\wedge(\partial_A\alpha)_d&=0,\\
\bar{\partial}\mu^{d}-\tfrac{1}{2}[\mu,\mu]^{d} & =0,\\ 
\partial\imath_{\mu}\Omega & =0,
\end{split}
\end{equation}
where we have redefined $b$ to be
\begin{equation}
b = (\mathcal{B}+ \ii \,\Delta \omega)_{(0,2)} - \mu^a \wedge x_a.
\end{equation}

Remarkably, the superpotential can still be written in a Chern--Simons fashion as
\begin{equation}
\Delta W=\int_{X}\langle y,\bar{D}y-\tfrac{1}{3}[y,y]-\partial b\rangle\wedge\Omega,\label{eq:gauge_action_with_b}
\end{equation}
where $y$ is now $(0,1)$-form valued in $\mathcal{Q}\simeq T^{*(1,0)}X\oplus\End V\oplus T^{(1,0)}X$, and $\bar{D}$, $[\cdot,\cdot]$ and $\langle\cdot,\cdot\rangle$ are now given by expressions (\ref{eq:Dbar_with_gauge}), (\ref{eq:bracket_with_gauge}) and (\ref{eq:pairing_with_gauge}). The corresponding equations of motion are
\begin{align}
\bar{D}y-\tfrac{1}{2}[y,y]-\tfrac{1}{2}\partial b & =0,\label{eq:y_eom_bundle}\\
\partial\imath_{\mu}\Omega & =0.\label{eq:b_eom_bundle}
\end{align}
Together with the vanishing of $\Delta W$, these are the conditions for a supersymmetric Minkowski solution.

\subsection{Vanishing of the superpotential}

We now want to understand the condition $\Delta W = 0$ in more detail. In what follows, we will consider the moduli problem with the gauge sector turned off. Everything we say goes through when we replace the $\bar{D}$ operator, bracket and pairing with those that include the gauge sector.

The deformed vacuum solution is supersymmetric if the equations of motion are satisfied and the superpotential itself vanishes. As $X$ is assumed to be compact, the superpotential vanishes if the terms wedged with $\Omega$ are $\bar\partial$-exact, that is
\begin{equation}
\langle y,\bar{D}y-\tfrac{1}{3}[y,y]-\partial b \rangle=\bar{\partial}\beta,
\end{equation}
where $\beta$ is an arbitrary $(0,2)$-form. Upon substituting the first equation of motion \eqref{eq:y_eom_bundle} into this expression, it simplifies to
\begin{equation}
\tfrac{1}{3!}\langle y,[y,y]\rangle-\tfrac{1}{2}y^{a}\partial_{a} b =\bar{\partial}\beta.\label{eq:vanishing}
\end{equation}
Now we use $\bar{D}^2 = 0$ to constrain $\beta$. Taking $\bar{D}$ of the first equation of motion \eqref{eq:y_eom_bundle} gives
\begin{equation}
\begin{split}0 & =\bar{D}^{2}y-[\bar{D}y,y]-\tfrac{1}{2}\bar{D}\partial b \\
 & \propto e^a(-[y,[y,y]]_a-[y,\partial b ]_a+\bar{\partial}\partial_{a} b) \\
 & =e^a\bigl(\tfrac{1}{3!}\partial_{a}\langle y,[y,y]\rangle-\tfrac{1}{2}\partial_{a}(y^{d}\partial_{d} b )+\bar{\partial}\partial_{a} b\bigr) \\
 & \propto\bigl(\partial\bar{\partial} b -\tfrac{1}{2}\partial(y^{d}\partial_{d} b )+\tfrac{1}{3!}\partial\langle y,[y,y]\rangle\bigr),
\end{split}
\end{equation}
where we have used $[y,[y,y]]_a=-\frac{1}{3!}\partial_{a}\langle y,[y,y]\rangle$ and $[y,\partial_{a}b]=\tfrac{1}{2}\partial_{a}(y^{d}\partial_{d}b)$.\footnote{These identities are easy to check using the explicit expressions for the bracket and pairing.} We can integrate this expression to give
\begin{equation}
k\,\bar\Omega=\bar{\partial} b -\tfrac{1}{2}y^{d}\partial_{d} b +\tfrac{1}{3!}\langle y,[y,y]\rangle,
\end{equation}
where $k$ is constant as it is anti-holomorphic and $X$ is compact. Combining this with the vanishing of the superpotential \eqref{eq:vanishing} gives
\begin{equation}
k\,\bar\Omega=\bar{\partial} b +\bar{\partial}\beta.
\end{equation}
As $\bar\Omega$ is not $\bar\partial$ exact, we must have $k=0$. We then identify $\beta=-b$ up to a $\bar\partial$-closed $(0,2)$-form. Putting this all together the full set of equations is
\begin{align}
\bar{D}y-\tfrac{1}{2}[y,y]-\tfrac{1}{2}\partial b  & =0,\label{eq:full_MC1}\\
\bar{\partial} b -\tfrac{1}{2}\langle y,\partial b\rangle +\tfrac{1}{3!}\langle y,[y,y]\rangle & =0,\label{eq:full_MC2}\\
\partial\imath_{\mu}\Omega & =0\label{eq:full_MC3}.
\end{align}
These equations are equivalent to the vanishing of the superpotential and its derivative, and so their solutions are a supersymmetric Minkowski vacuum. In other words, solutions $(y,b)$ to these equations are precisely the moduli of the Hull--Strominger system.

We pause to make a few comments. First note that these equations contain finitely many powers of $y$ and $b$; the equations do not give an infinite set of relations. This is somewhat striking -- generic deformations of geometric structures do not usually truncate at a given order. In our case the fact that the equations depend on terms up to $\mathcal{O}(y^3)$ is an indication that there is more structure to the Hull--Strominger system than at first sight. This extra structure is the existence of an underlying holomorphic Courant algebroid describing the subsector of deformations given by simultaneous deformations of the complex structure and the three-form flux. In deriving these equations and finding third-order equations for the moduli, we might be encouraged to think there is some sort of algebroid underlying the full heterotic system. Indeed, the ten-dimensional heterotic theory has a description in terms of generalised geometry~\cite{CMTW14,Garcia-Fernandez:2013gja,GRT15} and the Hull--Strominger system can be recast as in terms of holomorphic Courant algebroid~\cite{Garcia-Fernandez:2018emx}.

One might wonder if the form of these equations can survive $\alpha'$ corrections. We derived the equations for the moduli by starting from the superpotential for the four-dimensional theory that one would get by compactifying on a solution to the Hull--Strominger system. Part of the data of such solutions is a complex manifold. A complex manifold admits an $\SU3$ structure whose torsion is constrained~\cite{GH80}. A special case of such manifolds are those with vanishing torsion so they have $\SU3$ holonomy and are Calabi--Yau. If the solution to the Hull--Strominger system admits an $\alpha'\rightarrow 0$ limit, the $\alpha'=0$ solution is simply Calabi--Yau. In this case it is known that the superpotential receives no $\alpha'$ corrections (to finite order) and so, although the $\alpha'$-corrected geometry is not longer Calabi--Yau, the tree-level superpotential is exact~\cite{Witten86,Jardine:2018sft}. This means equations \eqref{eq:full_MC1}--\eqref{eq:full_MC3} will be correct even after $\alpha'$ corrections. It is not known what happens if there is no large-volume Calabi--Yau limit.

Up to this point there has been an asymmetry in the way we have treated the vanishing of $\Delta W$ and its derivative. In the next section we will see that we can combine these conditions into a single Maurer--Cartan equation for an $L_3$ algebra.

%%%%%%%%%%%%%%%%%%%%%%%%%%%%%%%%%%%%%%%%%%%%%%%%%%%%%%%%%%%%%%%%%%%%%%%%%%

\section{Moduli and an \texorpdfstring{$L_{3}$}{L3} algebra}\label{sec:L3}

So far we have derived the equations that determine the moduli for solutions to the Hull--Strominger system for finite deformations. We will show in this section that these equations can be reinterpreted as the Maurer--Cartan equation for an $L_3$ algebra. At first sight there is no obvious reason why the deformations of a system as complicated as the heterotic string should be described by such a ``nice'' algebra structure. However it is not as surprising if one remembers that the data of the Hull--Strominger system is equivalent to a holomorphic Courant algebroid with a holomorphic vector bundle~\cite{Garcia-Fernandez:2018emx}. The $L_\infty$ structures that govern deformations of Courant algebroids (or Dirac structures) have been found; in particular it is known that the deformation complex of a Dirac structure is isomorphic to a cubic $L_\infty$ or $L_3$ structure~\cite{Gualtieri:2017kdd,2012arXiv1202.2896F,2007JGP....57.1015K,2001math.....12152R}.

As we review in appendix \ref{L3_appendix}, an $L_{\infty}$ structure is specified by a choice of graded vector spaces $\mathcal{Y}_{n}$ and multilinear products $\ell_{k}$. The idea is that the conditions from the superpotential are most naturally written in terms of an $L_\infty$ structure that combines the action of $\bar D$ on the moduli fields $y$ and $\bar\partial$ on the $(0,2)$ moduli $b$. We take the vector spaces $\mathcal{Y}_{n}$ to be
\begin{equation}
\mathcal{Y}_{n}=\Omega^{(0,n)}(\Q)\oplus\Omega^{(0,n+1)}(X),
\end{equation}
so that an element of $\mathcal{Y}_{1}$ is $Y=(y,b)$ where $y$ is a $(0,1)$-form valued in $\Q$ and $b$ is a $(0,2)$-form. Using this notation, we write the multilinear products $\ell_{k}$ as
\begin{equation}
\label{eq:L3-brackets}
\begin{split}\ell_{1}(Y) & \coloneqq(\bar{D}y-\tfrac{1}{2}\partial b,\bar{\partial}b),\\
\ell_{2}(Y,Y) & \coloneqq([y,y],\langle y,\partial  b \rangle),\\
\ell_{3}(Y,Y,Y) & \coloneqq(0,-\langle y,[y,y]\rangle),\\
\ell_{k\geq4} & \coloneqq0.
\end{split}
\end{equation}
We give expressions for the $\ell_k$ where the entries are arbitrary elements of $\mathcal{Y}_n$ in \eqref{eq:Linfty_products}. One can check that these products have the correct symmetry properties and obey the $L_{\infty}$ relations, which we write in \eqref{eq:Linfty_relations}. Note that this is highly non-trivial and is an indication of the underlying holomorphic Courant algebroid.

\subsection{Quasi-isomorphism to a natural holomorphic \texorpdfstring{$L_{3}$}{L3} algebra}

We briefly remark that these structures have a nice mathematical interpretation: our $L_3$ algebra $(\mathcal{Y}, \ell_1, \ell_2, \ell_3)$ is $L_\infty$ equivalent to the underlying holomorphic algebra. 

Neglecting the gauge bundle for a moment, we have the sheaf $\mathcal{E}$ of $\bar D$-holomorphic sections of $\Q' \simeq T^{*(1,0)}X \oplus T^{(1,0)}X  $ and the sheaf of holomorphic functions $\OO_X$. These form an $L_3$ algebra, with underlying two-term complex
\begin{equation}
\label{eq:holo-L3}
	\OO_X \stackrel{\del}{\longrightarrow} \mathcal{E} ,%\Gamma(\Q) ,
\end{equation}
in precisely the same way that a Courant algebroid $E$ together with the real $\mathcal{C}^\infty$ functions form an $L_3$ algebra~\cite{Roytenberg-Weinstein} with two-term complex
\begin{equation}
	\mathcal{C}^\infty(\bbR) \stackrel{\dd}{\longrightarrow} E \simeq T^*X \oplus TX . %\Gamma(E) .
\end{equation}
One can then consider the Dolbeault resolutions of the sheaves $\mathcal{E}$ and $\OO_X$, and extend $\del$ to a morphism between them:
\begin{equation}
\label{eq:resolution}
\begin{tikzpicture}[scale=1.75,baseline=(current bounding box.center)] %% centers equation number
\node (Az) at (-2.5,1) {$0$}; 
\node (Ah) at (-1.3,1) {$\mathcal{O}_X$}; 
\node (A0) at (0,1) {$\mathcal{C}^\infty (\bbC)$}; 
\node (A1) at (1.5,1) {$\Omega^{(0,1)}$}; 
\node (A2) at (3.2,1) {$\Omega^{(0,2)}$}; 
\node (A3) at (4.5,1) {$$}; 
\node (Bz) at (-2.5,0) {$0$}; 
\node (Bh) at (-1.3,0) {$\mathcal{E}$}; 
\node (B0) at (0,0) {$\Q'$}; 
\node (B1) at (1.5,0) {$\Omega^{(0,1)}(\Q')$}; 
\node (B2) at (3.2,0) {$\Omega^{(0,2)}(\Q')$}; 
\node (B3) at (4.5,0) {$$}; 
\path[->,font=\scriptsize] 
(Az) edge node[above]{$$} (Ah)
(Ah) edge node[above]{$\iota$} (A0)
(A0) edge node[above]{$\bar\del$} (A1)
(A1) edge node[above]{$\bar\del$} (A2)
(A2) edge node[above]{$\bar\del$} (A3)
(Bz) edge node[above]{$$} (Bh)
(Bh) edge node[above]{$\iota$} (B0)
(B0) edge node[above]{$\bar{D}$} (B1)
(B1) edge node[above]{$\bar{D}$} (B2)
(B2) edge node[above]{$\bar{D}$} (B3)
(Ah) edge node[left]{$\del$} (Bh)
(A0) edge node[left]{$\del$} (B0)
(A1) edge node[left]{$\del$} (B1)
(A2) edge node[left]{$\del$} (B2);
\end{tikzpicture}
\end{equation}
Our complex $\mathcal{Y}$ is then the total complex of the deleted resolution and the differential $\ell_1$ is the natural differential on this (see e.g.~\cite{Weibel}). 
Our construction gives higher $\ell_n$ brackets on $\mathcal{Y}$, providing an $L_3$ algebra structure on the total complex. 

As~\eqref{eq:resolution} is simply a resolution of \eqref{eq:holo-L3}, 
this construction essential provides us with a reformulation of the holomorphic $L_3$ algebra~\eqref{eq:holo-L3} in terms of $\mathcal{C}^\infty$ objects. 
Explicitly, one has a map of complexes (of sheaves) as follows:
\begin{equation}
\label{eq:quasi-isomorphism}
\begin{tikzpicture}[scale=1.75,baseline=(current bounding box.center)] %% centers equation number
\node (Az) at (-2.5,1) {$0$}; 
\node (Ah) at (-1.6,1) {$\mathcal{O}_X$}; 
\node (A0) at (0,1) {$\mathcal{E}$}; 
\node (A1) at (2,1) {$0$}; 
\node (A2) at (4.3,1) {$0$}; 
\node (A3) at (5.7,1) {$$}; 
\node (Bz) at (-2.5,0) {$0$}; 
\node (Bh) at (-1.6,0) {$\mathcal{C}^\infty (\bbC)$}; 
\node (B0) at (0,0) {$\Gamma(E)\oplus \Omega^{(0,1)}$}; 
\node (B1) at (2,0) {$\Omega^{(0,1)}(E)\oplus\Omega^{(0,2)}$}; 
\node (B2) at (4.3,0) {$\Omega^{(0,2)}(E)\oplus\Omega^{(0,3)}$}; 
\node (B3) at (5.7,0) {$$}; 
\path[->,font=\scriptsize] 
(Az) edge node[above]{$$} (Ah)
(Ah) edge node[above]{$\del$} (A0)
(A0) edge node[above]{$$} (A1)
(A1) edge node[above]{$$} (A2)
(A2) edge node[above]{$$} (A3)
(Bz) edge node[above]{$$} (Bh)
(Bh) edge node[above]{$\ell_1$} (B0)
(B0) edge node[above]{$\ell_1$} (B1)
(B1) edge node[above]{$\ell_1$} (B2)
(B2) edge node[above]{$\ell_1$} (B3)
(Ah) edge node[left]{$\iota$} (Bh)
(A0) edge node[left]{$\iota$} (B0)
(A1) edge node[left]{$$} (B1)
(A2) edge node[left]{$$} (B2);
\end{tikzpicture}
\end{equation}
As the cohomology of the total complex is the same as the cohomology of the complex it is resolving, this is a quasi-isomorphism. (One can check this explicitly in our case.) 
However, the morphism in~\eqref{eq:quasi-isomorphism} also respects the bracket structure of the $L_3$ algebras on each complex, thus it is a quasi-isomorphism of $L_3$ algebras. We conclude that, in the $L_\infty$ sense, our $L_3$ algebra $(\mathcal{Y}, \ell_1, \ell_2, \ell_3)$ is equivalent to the holomorphic algebra~\eqref{eq:holo-L3}.

Including the gauge bundle in this construction is straightforward; one simply replaces the bundle $\Q'$ above with the full holomorphic bundle~\eqref{eq:holomorphic_bundle} (this also recently appeared in~\cite{Garcia-Fernandez:2018emx}). One finds an essentially identical two-term complex to~\eqref{eq:holo-L3}, giving an $L_3$ algebra on the local holomorphic sections. 
Via the Dolbeault resolution, one sees that this is quasi-isomorphic to our $L_3$ algebra $(\mathcal{Y}, \ell_1, \ell_2, \ell_3)$ (now including the gauge bundles) exactly as above.

%%%%%%%%%%%%%%%%%%%%%%%

\subsection{An \texorpdfstring{$L_{\infty}$}{L infinity} field equation}

As explained in \cite{Hohm:2017pnh}, there is a natural field equation that one can write down for a given $L_\infty$ structure. The constraint on the form of the field equation is that it is covariant under gauge transformations of the fields $\mathcal{Y}$. In terms of the $L_\infty$ products, the field equation is
\begin{equation}
\mathcal{F}(Y)  =\ell_{1}(Y)-\tfrac{1}{2}\ell_{2}(Y)-\tfrac{1}{3!}\ell_{3}(Y)+\ldots
\end{equation}
For us this expression truncates at third order as $\ell_{k\geq4}=0$.

Remarkably, the $L_{3}$ field equation coming from~\eqref{eq:L3-brackets} reproduces the conditions from the vanishing of the superpotential and its derivative:
\begin{equation}
\mathcal{F}(Y)  =(\bar{D}y-\tfrac{1}{2}\partial b -\tfrac{1}{2}[y,y],\bar{\partial} b -\tfrac{1}{2}\langle y, b \rangle+\tfrac{1}{3!}\langle y,[y,y]\rangle).
\end{equation}
In other words, the conditions for a supersymmetric Minkowski vacuum are equivalent to
\begin{equation}
\mathcal{F}(Y) = 0, \qquad \partial\imath_\mu \Omega=0.
\end{equation}

A particularly nice property of this rewriting is that the $L_{\infty}$ structure gives us the gauge transformations of the moduli for free and guarantees that the gauge algebra closes. The gauge transformation of $Y$ by a gauge parameter $\Lambda=(\lambda,\xi)\in\mathcal{Y}_{0}$ is
\begin{equation}\label{eq:L3_gauge_transforms}
\delta_{\Lambda}Y=\ell_{1}(\Lambda)+\ell_{2}(\Lambda,Y)-\tfrac{1}{2}\ell_{3}(\Lambda,Y,Y),
\end{equation}
where the higher-order brackets vanish. In general the gauge transformations take field equations to combinations of field equations -- the field equations are covariant. If one could construct an action that has the $L_3$ field equation as its equations of motion, one would expect that action to be invariant under the $L_3$ gauge transformations. In contrast to \cite{Hohm:2017pnh}, we have not been able to find such a candidate action nor do we expect one to exist; this is due to the fact that the supersymmetry conditions are the vanishing of the superpotential and its first derivative. Note that the superpotential alone is not expected to be invariant under both $\lambda$ and $\xi$ transformations -- we will see in the next section that the superpotential is actually invariant under the $\xi$ transformations alone which correspond to shifts by $\partial_a$-exact forms.

%%%%%%%%%%%%%%%%%%%%%%%%%%%%%%%%%%%%%%%%%%%%%%%%%%%%%%%%%%%%%%%%%%%%%%%%%%

\section{A reduced \texorpdfstring{$L_{3}$}{L3} algebra and an effective action}\label{sec:ReducedL3}

In this section we discuss some consequences of the $L_3$ algebra. In particular we comment on how the $L_3$ algebra can be reduced by quotienting by $\partial$-exact forms and show that this is equivalent to integrating out $b$. We also discuss the relation of the moduli $(y,b)$ to the effective theory one would find by compactifying on a solution to the Hull--Strominger system. 
As we have mentioned, the form of the superpotential \eqref{eq:gauge_action_with_b} closely resembles that of holomorphic Chern--Simons theory, and is in fact a generalisation of this theory. Holomorphic Chern--Simons theory has several interesting relations with mathematical disciplines such as open and closed topological string theory, knot theory, Donaldson--Thomas invariants and so on, and it would be interesting to look for heterotic generalisations of these relations. This will be the subject of future work. For now, we will restrict ourselves to making some observations about the (semi-) classical effective action \eqref{eq:gauge_action_with_b}, and its relation to the lower-dimensional effective physics. 

\subsection{Integrating out \texorpdfstring{$b$}{b}}

We want to integrate out the $(0,2)$-form field $b$. Looking back at the form of the $L_3$ gauge transformations \eqref{eq:L3_gauge_transforms}, taking $\Lambda=(0,\xi)$ gives
\begin{equation}
\delta y_a = \tfrac{1}{2} \partial_a \xi, \qquad \delta b = \bar\partial \xi - \tfrac{1}{2}\langle y, \partial \xi\rangle,
\end{equation}
for some $(0,1)$-form $\xi$. One can check that the superpotential \eqref{eq:gauge_action_with_b} is invariant under this gauge transformation provided $X$ is compact. From this we see that $y$ is defined up to $\partial_a$-exact forms. Notice also that $\Delta W$ splits into two pieces:
\begin{equation}\label{eq:superpotential_b}
\Delta W=\int_{X}\langle y,\bar{D}y-\tfrac{1}{3}[y,y]\rangle\wedge\Omega-\int_{X}\langle y,\partial b \rangle\wedge\Omega,
\end{equation}
where the second term can be written as
\begin{equation}
\int_{X}\langle y,\partial b\rangle\wedge\Omega\propto\int_{X}b\wedge\partial\imath_{\mu}\Omega.
\end{equation}
From this we see that $b$ plays the role of a Lagrange multiplier. We shall see below that given certain assumptions about the Hodge diamond of $X$, specifically $h^{(2,0)}=0$, then the field $b$ has no associated massless modes. We can then integrate out $b$, resulting in 
\begin{equation}
\label{eq:EffAction}
\Delta W[y]=\int_{X}\langle y,\bar{D}y-\tfrac{1}{3}[y,y]\rangle\wedge\Omega,
\end{equation} 
where $y$ now satisfies the constraint $\partial\imath_{\mu}\Omega=0$. We want to think of this functional as an effective action. Note that for $\partial\imath_{\mu}\Omega=0$ there is a gauge symmetry of this action
\begin{equation}
\delta y_{a}=\partial_{a}\xi,
\end{equation}
where $\xi\in\Omega^{0,1}(X)$. We are thus led to define $\tilde{Q}$ as a reduced sheaf of $\Q$ whose sections satisfy the constraint and are defined up to $\partial_a$-exact terms:
\begin{equation}
\Gamma(\tilde{Q}) = \{ \Gamma(Q) \,\,|\,\, \partial \imath_\mu \Omega = 0,\, y_a \sim y_a + \partial_a \xi \}.
\end{equation} 
One can check that the brackets on $\mathcal{Q}$ are well defined on $\tilde{\mathcal{Q}}$, and that the $L_3$ algebra descends to a differential graded Lie algebra (DGLA).

The second gauge symmetry of \eqref{eq:EffAction} is a generalisation of the Chern--Simons symmetry. The superpotential is invariant under
\begin{equation}
\label{eq:SymTrans}
\delta y=\bar{D}\lambda-[y,\lambda],
\end{equation}
where $\lambda\in\Omega^0(\tilde \Q)$ satisfies $\partial\imath_\lambda \Omega =0$. Note that the gauge algebra generated by \eqref{eq:L3_gauge_transforms} is reducible; a gauge transformation by $\Lambda=(\lambda,\xi)$ is trivial if
\begin{equation}
\lambda = \partial w,\qquad \xi = -\bar\partial w + \tfrac12 \langle y,\partial w\rangle,
\end{equation}
for $w\in\Omega^{0}(X)$.

From the effective superpotential \eqref{eq:EffAction} we derive the equation of motion
\begin{equation}
\label{eq:MCfull}
\bar D y-\tfrac12[y,y]= 0.
\end{equation}
This should be interpreted as an equation on the sheaf $\tilde \Q$. Note that under $y_a \mapsto y_a + \partial_a \xi$, this equation becomes
\begin{equation}
\label{eq:MCfull2}
\bar D y-\tfrac12[y,y]= \partial (\bar\partial \xi + y^d \partial_d \xi)\sim 0,
\end{equation}
so it is well defined as an equation on the sheaf $\tilde \Q$. Recall that we already know the superpotential is invariant under $\delta y_a = \partial_a\xi$. One can show that equation \eqref{eq:MCfull}, together with the condition that the effective action vanishes, is equivalent to the Maurer--Cartan equations \eqref{eq:full_MC1}--\eqref{eq:full_MC3}. Indeed, note that \eqref{eq:MCfull} is equivalent to
\begin{equation}
\bar D y-\tfrac12[y,y]=\tfrac12 \p  b,
\end{equation}
for some $b\in\Omega^{(0,2)}(X)$. For solutions to this equation, the condition that the action vanishes can then be written as $\langle y,[y,y]\rangle=\bp\textrm{-exact}$, which can be rewritten as
\begin{equation}
\tfrac{1}{3!}\langle y,[y,y]\rangle-\tfrac{1}{2}y^{a}\partial_{a} b =\bar{\partial}\beta,
\end{equation}
for some $\beta\in\Omega^{(0,2)}(X)$. Here we used that the second term on the left-hand side is $\bar\partial$-exact -- it integrates to zero against $\Omega$ due to the constraint $\partial\imath_{\mu}\Omega=0$. This then gives the same starting point for our derivation of equations \eqref{eq:full_MC1}--\eqref{eq:full_MC3}. 

It is beyond the scope of the present paper to investigate general solutions to \eqref{eq:MCfull} and $\Delta W[y]=0$, i.e.~integrable deformations of heterotic geometries. We will however make some comments on the couplings derived from \eqref{eq:EffAction} in the four-dimensional effective field theory. From this we make a conjecture about the  obstructions that can appear in the Maurer--Cartan equations. 

\subsection{Effective field theory and Yukawa couplings}

Our starting point to derive the effective physics is the superpotential \eqref{eq:gauge_action_with_b}, where we have re-introduced the field $b$. When dimensionally reducing the theory, it is common practice to split our fields $(y,b)$ into ``massless" and ``heavy" modes
\begin{align}
y&=y_0+y_\text{h},\\
b&=b_0+b_\text{h}.
\end{align}
We imagine performing a formal dimensional reduction of the theory to a four-dimensional Minkowski background where we keep all the massive Kaluza--Klein modes for the time being. The corresponding mass matrix of the reduced theory reads\footnote{In principle, there is also a potential coming from $D$-terms. However, as we show in appendix \ref{app:Dterms}, the $D$-terms can be set to zero by a complexified gauge transformation and so they do not lift any moduli.}
\begin{equation}
V_{\alpha\bar\beta}=e^{\cal K}\p_{\alpha}\p_{\gamma}W\p_{\bar \beta}\p_{\bar\kappa}\bar W\,{\cal K}^{\gamma\bar\kappa},
\end{equation}
where $\{\alpha,\beta,\ldots\}$ denote holomorphic directions in the parameter space and $\cal K$ is the K\"ahler potential. Full knowledge of the K\"ahler potential is not necessary at this point, but the curious reader is referred to \cite{COM17,McOrist16} for more details. From the form of the mass matrix, it is easy to see that a field direction $\alpha$ is massless if and only if
\begin{equation}
\label{eq:Wmassless}
\p_{\gamma}\p_{\alpha}W=0\quad\forall\;\gamma\quad\Rightarrow\quad (\p_{\gamma}\p_{\alpha}\Delta W)\vert_{(y,b)=0}=0\quad\forall\;\gamma,
\end{equation}
where the field directions $\gamma$ can in principle be massive.

In the end, we are interested in a reduced field theory of massless modes, where the massive modes have been ``integrated out''. It is easy to see that the field direction corresponding to $b$ is massive (although need not be an eigenmode of the mass matrix). Indeed, from \eqref{eq:Wmassless} it follows that $b_0$ must satisfy
\begin{equation}
\p_a b_0=0,
\end{equation}
and so $b$ is an anti-holomorphic section of $\Omega^{(0,2)}(X)$. We restrict ourselves to geometries where this bundle has no sections, in other words
\begin{equation}
\HH^{(2,0)}_{\bp}(X)=0.
\end{equation}
This is true in particular for Calabi--Yau geometries. It follows that we can integrate out the $b$ field as far as the effective theory is concerned, leaving us with the effective superpotential \eqref{eq:EffAction}, where now the Beltrami differential component $\mu$ of $y$ satisfies $\partial\imath_{\mu}\Omega=0$ as above. From condition \eqref{eq:Wmassless} it follows that the remaining massless fields $y_0$ then satisfy
\begin{equation}
\label{eq:MasslessFull}
\bar Dy_0=0,
\end{equation}
where this should be viewed as an equation in the sheaf $\tilde \Q$. 

It is also natural to decompose the symmetry transformations \eqref{eq:SymTrans} in terms of the massless and massive modes. A suggestive decomposition, given the condition \eqref{eq:MasslessFull}, is the following
\begin{align}
\delta y_0&=\bar D\lambda,\\
\delta y_\text{h}&=-[y_0,\lambda]-[y_\text{h},\lambda].
\end{align}
With this decomposition, we see that the massless modes are parametrised by cohomology classes
\begin{equation}
[y_0]\in \HH^{(0,1)}_{\bar D}(\tilde \Q).
\end{equation}
This cohomology is isomorphic to $\HH^{(0,1)}_{\bar D}(\Q)$ for manifolds satisfying either  the $\p\bp$-lemma or $\HH^{(0,1)}(X)=0$ \cite{AGS14, OS14b}. We give a brief review of this cohomology and its decomposition in into more familiar cohomologies by means of long exact sequences in appendix \ref{app:massless}.

Decomposed in terms of massless and heavy modes, the effective action now reads
\begin{equation}
\Delta W=\int_X\bigl(\langle y_\text{h},\bar D_{y_0}y_\text{h}\rangle-\tfrac13\langle y_0,[y_0,y_0]\rangle-\langle y_\text{h},[y_0,y_0]\rangle-\tfrac13\langle y_\text{h},[y_\text{h},y_\text{h}]\rangle\bigr)\wedge\Omega,
\end{equation}
where we denote
\begin{equation}
\bar D_{y_0}y_\text{h}=\bar Dy_\text{h}-[y_0,y_\text{h}].
\end{equation}
We see that the heavy $y_\text{h}$ modes are the only ones that propagate internally.
Note that even though we take the expectation value of $y_\text{h}$ to vanish, by including internal quantum corrections, we see that the coupling between $y_0$ and $y_\text{h}$ can generate higher-order couplings of the massless fields. These new couplings are however of quartic order and higher in $y_0$, and are hence non-renormalisable in the effective field theory. The only renormalisable coupling we need to worry about from an effective field theory point of view is therefore the Yukawa coupling
\begin{equation}
\label{eq:Yukawa}
\Delta W_{\text{Yuk}}(y_0)=-\tfrac13\int_X\langle y_0,[y_0,y_0]\rangle\wedge\Omega.
\end{equation}
This argument is similar to and generalises Witten's standard argument for the gauge sector~\cite{Witten86}. 

Note that in addition to the standard Yukawa couplings between bundle moduli, the Yukawa couplings \eqref{eq:Yukawa} also contain couplings of gravity-gravity type (couplings of deformations of the geometry) and gravity-bundle type, often referred to as $\mu$-terms in the literature. It would be interesting to investigate the phenomenological implications of such couplings, but it is beyond the scope of the present paper to do so. 

Let $\alpha^A\in \HH^{(0,1)}(\tilde \Q)$ denote a set of inequivalent cohomology classes spanning $\HH^{(0,1)}(\tilde \Q)$, and expand
\begin{equation}
y_0=\sum_AC_A\alpha^A,
\end{equation}
where the $C_A$ now correspond to the four-dimensional fields, including in principle moduli and matter fields. The Yukawa couplings then read
\begin{equation}
\Delta W_{\text{Yuk}}=\sum_{A,B,C}C_AC_BC_C\int_X\langle \alpha^A,[\alpha^B,\alpha^C]\rangle\wedge\Omega.
\end{equation}
A massless field direction $\alpha^A$ is then truly free if and only if 
\begin{equation}
Y_{ABC}=\int_X\langle \alpha^A,[\alpha^B,\alpha^C]\rangle\wedge\Omega=0\quad\forall\;\alpha^B,\alpha^C.
\end{equation}
In particular, this is true if 
\begin{equation}
\label{eq:MC2}
[\alpha^A,\alpha^B]=\bar D\beta^{AB}\quad\forall\;\alpha^B.
\end{equation}
Note that, starting from the Maurer--Cartan equation \eqref{eq:MCfull}, this is simply the condition for an infinitesimal deformation in the field direction $\alpha^A$ to be unobstructed. The effective field theory then prompts us to make the following conjecture: when $\HH^{(2,0)}_{\bp}(X)=0$,\footnote{So we can integrate out $b$ in the effective theory.} the only non-trivial obstructions coming from the Maurer--Cartan equations are given by the constraints \eqref{eq:MC2} on the infinitesimal moduli.

%%%%%%%%%%%%%%%%%%%%%%%%%%%%%%%%%%%%%%%%%%%%%%%%%%%%%%%%%%%%%%%%%%%%%%%%%%

\section{Conclusions}\label{sec:Con}

In this paper we have considered finite deformations of the Hull--Strominger system. Starting with the four-dimensional $\mathcal{N}=1$ superpotential, we showed that integrable deformations corresponding to holomorphic directions on the moduli space can be parametrised by solutions of a Maurer--Cartan equation for an $L_3$ algebra, which we described in detail.

There are many directions one could follow from this work. Firstly, one might wonder which of the infinitesimal deformations parametrised by $\HH^{(0,1)}_{\bar D}(\tilde Q)$ can be integrated to finite deformations, corresponding to solutions of the $L_3$ Maurer--Cartan equation. In particular, are there some special cases where a generalisation of the Tian--Todorov lemma applies? It would be interesting to apply our formalism to some explicit examples, and in this way work out the spectrum of free fields in the low-energy four-dimensional theory.

The superpotential led to a generalisation of holomorphic Chern--Simons theory~\cite{Thomas97, DT98, Thomas00} that couples hermitian and complex structure moduli. Following \cite{Thomas97} and \cite{BCO+94}, it seems that one should think of this theory by taking spacetime to be spanned by the anti-holomorphic directions with holomorphic $\Omega$-preserving generalised diffeomorphisms playing the role of a gauge group.\footnote{Generalised diffeomorphisms are transformations generated by sections of $Q$ via the Courant bracket. The $\Omega$-preserving condition is just \eqref{eq:constraint} which ensures the deformed three-form is $\dd$-closed.} It would be interesting to study this further. In particular, by starting from the superpotential as an effective action and investigating its quantisation one might hope it defines a consistent quantum theory (cf.~\cite{Giusto:2012jm,Costello:2015xsa}) and gives analogues of Donaldson--Thomas or holomorphic Casson invariants for heterotic geometries. Note that the superpotential is complex in general so the path integral will not be convergent. Such complex path integrals have appeared before in the study of complex Chern--Simons theory~\cite{Witten:1989ip,Witten:2010cx,Witten:2010zr} where they are understood by analytic continuation. We foresee a similar treatment here.

As a step towards a complete understanding of the quantum heterotic moduli space, one could construct a world-sheet AKSZ topological model \cite{Alexandrov:1995kv, Ikeda:2012pv} or a topological string model for the effective theory similar to Witten's open string model for ordinary Chern--Simons theory~\cite{Witten:1992fb}. As a guide, one might start by comparing the heterotic moduli space with the spectrum of holomorphic $\beta\gamma$ systems and the chiral de Rham complex~\cite{Witten07, Kapustin05, Nekrasov05, Zeitlin08,2016arXiv161009657G,Gwilliam:2017axm}. Several other approaches to the $(0,2)$ world-sheet have appeared over the years (see \cite{Adams:2005tc, Zeitlin06, Mason:2007zv, McOrist:2010ae,MQS+13, Zeitlin15, Casali:2015vta, Gwilliam:2017axm, Jardine:2018sft} and references therein). It would be interesting to investigate how these methods connect with the approach outlined in the present paper. These are all interesting aspects which we hope to explore in future publications. 

One might also consider the moduli space of heterotic compactifications on more exotic geometries, such as $\text{G}_2$ or $\text{Spin}(7)$ manifolds~\cite{Clarke:2016qtg, delaOssa:2017pqy, Fiset:2017auc}. In the case of $\text{G}_2$ compactifications, the form of the moduli space is remarkably similar: for example, the infinitesimal deformations are again captured by a cohomology. Despite this there are notable differences such as the analogue of the bundle $\mathcal{Q}$ not appearing as an extension. It would be interesting to investigate the finite deformation algebras in these cases, and in the process identify the corresponding $L_\infty$ structure. This might give a $\text{G}_2$ generalisation of Chern--Simons theory.

We have been concerned with finding the honest supersymmetric deformations of solutions to the Hull--Strominger system. For this we only needed to consider the superpotential in the four-dimensional theory. Of course, the four-dimensional theory also has a K\"ahler potential which is important for understanding the physical potential of the effective theory. The K\"ahler potential and the metric on the moduli space have been worked out in recent publications~\cite{COM17,McOrist16}. One might wonder how these objects appear in our formalism. It seems that the cleanest description of these objects would follow from a proper analysis using generalised geometry. As outlined in appendix~\ref{app:N=1}, the $\mathcal{N}=1$ heterotic structure is described by an invariant object $\tilde\chi$ so the K\"ahler potential should be given by a functional of this object, similar to the Hitchin functional for $\SU3$ and $\text{G}_2$ structures~\cite{Hitchin01}. Indeed, it seems that a similar story applies to heterotic compactifications in other dimensions. We hope to make progress on this in a future work.

Note that even though we have the invariant object $\tilde\chi$ we do not have a natural integrability condition for it -- the generalised connection is not torsion-free in the heterotic string~\cite{CMTW14}. Curiously, it appears that when one looks at deformations of this structure there is a nice integrability condition (given by the superpotential). It would be interesting to see if this pattern persists for other generalised geometries.

\subsection*{Acknowledgements}

We would like to thank Bobby Acharya, Dmitri Alekseevsky, Philip Candelas, Jos\'e Figueroa-O'Farrill, Marco Gualtieri, Brent Pym, Savdeep Sethi and Dan Waldram for helpful discussions. AA is supported by a Junior Research Fellowship from Merton College, Oxford. XD is supported in part by the EPSRC grant EP/J010790/1. CS-C has been supported by a Seggie Brown Fellowship from the University of Edinburgh. ESS is supported by a grant from the Simons Foundation (\#488569, Bobby Acharya).

\appendix

\section{Conventions}

In this appendix we set out our conventions and notation. We will use $(m,n,\ldots)$ indices to denote real coordinates and $(a,b,\ldots,\bar{a},\bar{b},\ldots)$ to denote complex coordinates on the real six-dimensional manifold $X$. Using this we can expand, for example, a vector as
\begin{equation}
v=v^{m}\hat{e}_{m}=v^{a}\hat{e}_{a}+v^{\bar{a}}\hat{e}_{\bar{a}}.
\end{equation}
Our fields are form-valued so, to save space, we often omit the wedge symbol where it will not lead to confusion.
Our convention for the contraction of a vector-valued one-form with a $p$-form is that the vector index is contracted as usual and the form components are wedged. In coordinates, for a vector-valued one-form $w$ and a $p$-form $\rho$, we have
\begin{equation}
\imath_w \rho = e^m \wedge \imath_{w_m} \rho = w^n \wedge \rho_n,
\end{equation}
where $\rho_m$ is defined as
\begin{equation}
\rho_m = \frac{1}{(p-1)!} \rho_{m n_1 \ldots n_{p-1}}e^{n_1\ldots n_{p-1}}.
\end{equation}
It follows that $\imath_w$ satisfies a Leibniz rule:
\begin{equation}
\imath_w (\rho \wedge \sigma) = \imath_w \rho \wedge \sigma + \rho \wedge \imath_w \sigma.
\end{equation}
The interior product of a vector with a one-form is extended to $p$-vectors and $p$-forms using the $\lrcorner$ operation, defined as
\begin{equation}
u \lrcorner \rho = \tfrac{1}{p!}u^{m_1 \ldots m_p}\rho_{m_1\ldots m_p},
\end{equation}
where $u$ is a $p$-vector and $\rho$ is a $p$-form. We indicate the $p$-vector obtained by raising the indices of a $p$-form with the metric $g$ by a superscript $\sharp$ -- for example
\begin{equation}
(\rho^\sharp)^{m_1 \ldots m_p} = g^{m_1 n_1}\ldots g^{m_p n_p}\rho_{n_1 \ldots n_p}.
\end{equation}

\subsection{Heterotic supergravity}

The Hull--Strominger system follows from setting the supersymmetry variations of the ten-dimensional gravitino $\psi$, dilatino $\lambda$ and gaugino $\chi$ to zero. In our conventions these are
\begin{align}
\delta\psi_M &= \nabla_M^+ \varepsilon = \nabla^\text{LC}_M\varepsilon+\tfrac{1}{8}H_{MNP}\Gamma^{NP}\varepsilon + \mathcal{O}(\alpha'^2),\label{eq:SUSY1}\\ 
\delta \lambda &= (\Gamma^M \partial_M \phi + \tfrac{1}{12} H_{MNP}\Gamma^{MNP})\varepsilon + \mathcal{O}(\alpha'^2),\\
\delta \chi &= -\tfrac12 F_{MN}\Gamma^{MN}\varepsilon + \mathcal{O}(\alpha'^2),\label{eq:SUSY3}
\end{align}
where $\varepsilon$ is a ten-dimensional Majorana--Weyl spinor and $\nabla^\text{LC}$ denotes the Levi-Civita connection.

\subsection{Holomorphic structure}

Ignoring the gauge sector for the moment, the relevant fields are $(0,1)$-forms taking values in $\mathcal{Q}'=T^{*(1,0)}X\oplus T^{(1,0)}X$. We write $(0,1)$-forms valued in this bundle as
\begin{equation}
y=(x_{a}e^{a},\mu^{a}\hat{e}_{a}) = x + \mu ,
\end{equation}
where $x_{a}$ and $\mu^{a}$ are $(0,1)$-forms. More generally we will write $y_{p}=(x_{p},\mu_{p})\in\Omega^{(0,p)}(\Q')$ -- we will often drop the subscript denoting the form degree. 

The holomorphic structure is defined by a $\bar{D}$ operator that is nilpotent. The action of $\bar{D}$ on $y\in\Omega^{(0,p)}(\Q')$ is
\begin{equation}
\label{eq:Dbar_definition}
\begin{split}(\bar{D}y)_{a} & =\bar{\partial}x_{a}+\ii(\partial\omega)_{ea\bar{c}}e^{\bar{c}}\wedge\mu^{e},\\
(\bar{D}y)^{a} & =\bar{\partial}\mu^{a}.
\end{split}
\end{equation}
One can check that $\bar{D}^{2}=0$ follows from $\dd H\propto\partial\bar{\partial}\omega=0$. Note also that this convention implies
\begin{equation}
(\bar D \partial \rho)_a = \bar\partial(\partial_a \rho) = \partial_a \bar\partial \rho,
\end{equation}
for a form $\rho$ (not valued in $\mathcal{Q}'$).

The bracket $[\cdot,\cdot]\colon\Omega^{(0,p)}(\Q')\times\Omega^{(0,q)}(\Q')\rightarrow\Omega^{(0,p+q)}(\Q')$ is
\begin{equation}
\begin{split}[y,y']_{a} & =\mu^{d}\wedge\partial_{d}x'_{a}-\partial_{d}x_{a}\wedge\mu'^{d}-\tfrac{1}{2}\mu^{d}\wedge\partial_{a}x'_{d}+\tfrac{1}{2}\partial_{a}x_{d}\wedge\mu'^{d}+\tfrac{1}{2}\partial_{a}\mu^{d}\wedge x'_{d}-\tfrac{1}{2}x_{d}\wedge\partial_{a}\mu'^{d},\\{}
[y,y']^{a} & =\mu^{b}\wedge\partial_{b}\mu'^{a}-\partial_{b}\mu^{a}\wedge\mu'^{b}.
\end{split}
\label{eq:bracket_definintion}
\end{equation}
The bracket is graded commutative and satisfies $[y_{p},y_{q}]=(-)^{1+pq}[y_{q},y_{p}]$. Furthermore, $\bar{D}$ satisfies a graded Leibniz identity with the bracket
\begin{equation}
\bar{D}[y_{p},y'_{q}]=[\bar{D}y_{p},y'_{q}]+(-1)^{p}[y_{p},\bar{D}y'_{q}].
\end{equation}
The bracket does not satisfy a graded Jacobi identity. As is the case for a Courant algebroid, the Jacobi identity holds up to a $\partial$-exact term. Evaluated for $y\in\Omega^{(0,1)}(\Q')$, one finds
\begin{equation}
[y,[y,y]]=-\frac{1}{3!}\partial\langle y,[y,y]\rangle,
\end{equation}
so that only the $(1,0)$-form valued component is non-zero. Here we have defined a pairing between two sections as
\begin{equation}
\langle y,y'\rangle=\mu^{d}\wedge x'_{d}+x_{d}\wedge\mu'^{d}.\label{eq:pairing_definition}
\end{equation}
More generally one finds
\begin{equation}
\begin{split} & [y_{m},[y'_{n},y''_{p}]]+(-1)^{m(n+p)}[y'_{n},[y''_{p},y_{m}]]+(-1)^{p(m+n)}[y''_{p},[y_{m},y'_{n}]]\\
 & =-\tfrac{1}{3!}\Bigl[\partial_{a}\langle y,[y',y'']\rangle+(-1)^{m(n+p)}\partial_{a}\langle y',[y'',y]\rangle+(-1)^{p(m+n)}\partial_{a}\langle y'',[y,y']\rangle\Bigr].
\end{split}
\end{equation}

When we include the gauge sector we write sections as
\begin{equation}
y=(x_{a}e^{a},\alpha,\mu^{a}\hat{e}_{a}) = x + \alpha+ \mu \in\Omega^{(0,1)}(\Q),
\end{equation}
where $x_{a}$ and $\mu^{a}$ are $(0,1)$-forms. The $\bar{D}$ operator, bracket and pairing are extended to include the gauge field component. The $\bar{D}$ operator is
\begin{equation}
\begin{split}(\bar{D}y)_{a} & =\bar{\partial}x_{a}+\ii(\partial\omega)_{ea\bar{c}}e^{\bar{c}}\wedge\mu^{e}-\tr(F_{a}\wedge\alpha),\\
(\bar{D}y)^{a} & =\bar{\partial}\mu^{a},\\
(\bar{D}y)_{\alpha} & =\bar{\partial}_{A}\alpha+F_{b}\wedge\mu^{b},
\end{split}
\label{eq:Dbar_with_gauge}
\end{equation}
where the final component is the gauge field piece and $F_{a}=F_{a\bar{b}}\dd z^{\bar{b}}$. The bracket is
\begin{equation}
\begin{split}[y,y']_{a} & =\ldots-\tfrac{1}{2}\tr\alpha\wedge(\partial_A\alpha')_a+\tfrac{1}{2}\tr(\partial_A\alpha)_a\wedge\alpha',\\
[y,y']^{a} & =\ldots,\\{}
[y,y']_{\alpha} & =-\alpha\wedge\alpha'-(-)^{1+\alpha\alpha'}\alpha'\wedge\alpha+\mu^{b}\wedge(\partial_A\alpha')_b-(\partial_A\alpha)_b\wedge\mu'^{b}\\
 & =-[\alpha,\alpha']+\mu^{b}\wedge(\partial_A\alpha')_b-(\partial_A\alpha)_b\wedge\mu'^{b},
\end{split}
\label{eq:bracket_with_gauge}
\end{equation}
where we have written only the extra terms that appear in the bracket. Again, the bracket obeys a Jacobi identity up to a $\partial$-exact term. The pairing between sections is given by
\begin{equation}
\langle y,y'\rangle=\mu^{d}\wedge x'_{d}+x_{d}\wedge\mu'^{d}+\tr(\alpha\wedge\alpha').\label{eq:pairing_with_gauge}
\end{equation}

\subsection{\texorpdfstring{$L_{\infty}$}{L infinity} structure}\label{L3_appendix}

We follow \cite{Hohm:2017pnh} for the conventions of an $L_{\infty}$ algebra in the ``$\ell$-picture''.\footnote{We have swapped $n\rightarrow -n$ so that the degree of $\mathcal{Y}_n$ matches the form degree of $y$} We start with a graded vector space $\mathcal{Y}$
\begin{equation}
\mathcal{Y}=\bigoplus_{n}\mathcal{Y}_{n},\qquad n\in\mathbb{Z},
\end{equation}
where the $\mathcal{Y}_{n}$ are of degree $n$. The $L_{\infty}$ algebra admits multilinear products $\ell_{1}$, $\ell_{2}$, $\ldots$, where $\ell_{k}$ has degree $2-k$. This means $\ell_{1}$ is degree $1$, $\ell_{2}$ is degree $0$, $\ell_{3}$ is degree $-1$, and so on. The products are graded commutative so that for example
\begin{equation}
\ell_{2}(Y,Y')=(-1)^{1+YY'}\ell_{2}(Y',Y),
\end{equation}
where a superscript denotes the degree of $Y\in\mathcal{Y}_{n}$. More generally we have
\begin{equation}
\ell_{k}(Y^{\sigma(1)},\ldots,Y^{\sigma(k)})=(-1)^{\sigma}\epsilon(\sigma;Y)\ell_{k}(Y^{1},\ldots,Y^{k}),
\end{equation}
where $Y^{1}=Y$, $Y^{2}=Y'$, etc. The sign has two contributions: $(-1)^{\sigma}$ gives a plus if the permutation is even and a minus if the permutation is odd; $\epsilon(\sigma;Y)$ is determined by
\begin{equation}
Y^{1}\wedge\ldots\wedge Y^{k}=\epsilon(\sigma;Y)Y^{\sigma(1)}\wedge\ldots\wedge Y^{\sigma(k)},
\end{equation}
where $Y\wedge Y^{'}=(-1)^{YY'}Y'\wedge Y$.

In these conventions, the first few $L_{\infty}$ identities are
\begin{equation}\label{eq:Linfty_relations}
\begin{split}\ell_{1}(\ell_{1}(Y)) & =0,\\
\ell_{1}(\ell_{2}(Y,Y')) & =\ell_{2}(\ell_{1}(Y),Y')+(-1)^{Y}\ell_{2}(Y,\ell_{1}(Y')),\\
\ell_{1}(\ell_{3}(Y,Y',Y'')) & =-\ell_{3}(\ell_{1}(Y),Y',Y'')-(-1)^{Y}\ell_{3}(Y,\ell_{1}(Y'),Y'')\\
 & \eqspace -(-1)^{Y+Y'}\ell_{3}(Y,Y',\ell_{1}(Y''))  -\ell_{2}(\ell_{2}(Y,Y'),Y'')\\
  & \eqspace-(-1)^{(Y+Y')Y''}\ell_{2}(\ell_{2}(Y'',Y),Y')-(-1)^{(Y'+Y'')Y}\ell_{2}(\ell_{2}(Y',Y''),Y)
\end{split}
\end{equation}
The field equations and gauge transformations are
\begin{align}
\mathcal{F}(Y) & =\sum_{n=1}^{\infty}\frac{(-1)^{n(n-1)/2}}{(n)!}\ell_{n}(Y^{n})=\ell_{1}(Y)-\tfrac{1}{2}\ell_{2}(Y,Y)-\tfrac{1}{3!}\ell_{3}(Y,Y,Y)+\ldots,\\
\delta_{\Lambda}Y & =\ell_{1}(\Lambda)+\ell_{2}(\Lambda,Y)-\tfrac{1}{2}\ell_{3}(\Lambda,Y,Y)-\tfrac{1}{3!}\ell_{4}(\Lambda,Y,Y,Y)+\ldots,
\end{align}
where $Y\in\mathcal{Y}_{1}$ and $\Lambda\in\mathcal{Y}_{0}$.

The multilinear products $\ell_k$ for the moduli of the heterotic system are
\begin{equation}\label{eq:Linfty_products}
\begin{split}\ell_{1}(Y) & \coloneqq(\bar{D}y+\tfrac{1}{2}(-1)^Y\partial b,\bar{\partial}b),\\
\ell_{2}(Y,Y') & \coloneqq([y,y'],\tfrac{1}{2}(\langle y,\partial b'\rangle+(-1)^{1+YY'}\langle y',\partial b\rangle)),\\
\ell_{3}(Y,Y',Y'') & \coloneqq\tfrac{1}{3}(-1)^{Y+Y'+Y''}(0,\langle y,[y',y'']\rangle+(-1)^{Y(Y'+Y'')}\langle y',[y'',y]\rangle\\
& \phantom{\coloneqq\tfrac{1}{3}(-1)^{Y+Y'+Y''}(0,}+(-1)^{Y''(Y+Y')\,}\langle y'',[y,y']\rangle),\\
\ell_{k\geq4} & \coloneqq 0.
\end{split}
\end{equation}

%%%%%%%%%%%%%%%%%%%%%%%%%%%%%%%%%%%%%%%%%%%%%%%%%%%%%%%%%%%%%%%%%%%%%%%%%%

\section{Comments on heterotic flux quantisation}
\label{sec:flux_quant}

In this appendix we comment on flux quantisation in the heterotic string, and its relation to the global nature of the deformation of certain quantities that appear in this paper. Our discussion applies to general ten-dimensional heterotic supergravity and can thus be applied to solutions other than four-dimensional Minkowski compactifications. Note that we are not saying anything new here; understanding flux quantisation in the heterotic string is still an open problem for the simple reason that $H$ is not $\dd$-closed in general.

We proceed in steps, beginning with a toy example of the quantisation of the flux of an abelian gauge bundle. We then present a similar treatment of the two-form gerbe as a warm-up for the case of the heterotic gauge fields. We follow~\cite{Hitchin:1999fh} for much of the early part of the discussion.

\subsection{A toy example: The abelian bundle}

Consider a vector bundle over a manifold $X$ and let $A$ denote an abelian connection with curvature $F=\d A$. Let $\{{\cal U}^i\}$ denote an open cover of $X$. We denote the overlaps by ${\cal U}^{ij}={\cal U}^i\cap{\cal U}^j$ and so on for higher intersections. 
We assume that the cover is ``good'' so that the ${\cal U}^i$ and their intersections are contractible. We employ the standard notion for the \v{C}ech co-boundary operator where appropriate: for some sheaf $\mathcal{F}$, if $f_i \in \mathcal{F}(\cU^i)$ then $(\dch f)_{ij} = f_i - f_j \in \mathcal{F}(\cU^{ij})$, and so on.

For the curvature to be well defined, we require that on ${\cal U}^{ij}$ we have
\begin{equation}
\d(A_i-A_j)=0,
\end{equation}
where $A_i$ and $A_j$ denotes the connections on ${\cal U}^{i}$ and ${\cal U}^{j}$. As ${\cal U}^{ij}$ is contractible, by the Poincar\'e lemma, we must therefore have
\begin{equation}
\label{eq:patchA}
(\dch A)_{ij} = A_i-A_j=\d\gamma_{ij},
\end{equation}
on ${\cal U}^{ij}$ for some zero-forms $\gamma_{ij}$. On triple overlaps ${\cal U}^{ijk}$ we have
\begin{equation}
	\d (\dch\gamma)_{ijk} = 
	\d(\gamma_{ij}+\gamma_{jk}+\gamma_{ki})=(\dch^2 A)_{ijk}=0,
\end{equation}
which is often referred to as taking the co-cycle of $\d\gamma_{ij}$. It follows that $c_{ijk} =(\dch \gamma)_{ijk}$ are constants: 
\begin{equation}
\label{eq:Const1}
c_{ijk}=(\dch \gamma)_{ijk} = \gamma_{ij}+\gamma_{jk}+\gamma_{ki} 
	\,\,\in\,\, \bbR(\cU^{ijk}) \subset \cC^{\infty}(\cU^{ijk};\bbR).
\end{equation}
Clearly from~\eqref{eq:Const1} we have $\dch c = 0$, so the $c_{ijk}$ define a class of the sheaf cohomology
\begin{equation}
[c_{ijk}]\in\check \HH{}^2(X;\bbR)\cong \HH^2(X;\bbR),
\end{equation}
which represents the two-form flux $F=\d A$. From \eqref{eq:Const1} this might look like a trivial co-cycle, but this is deceptive since the class is trivial only if the individual $\gamma_{ij}$ can be chosen to be constant. In this case, we see from \eqref{eq:patchA} that the connection $A$ can be made global so that the flux is trivial. In this language flux quantisation is the statement that the $c_{ijk}$ are in fact integers, $c_{ijk} \in \bbZ(\cU^{ijk})$, so that they define an integral class $[c_{ijk}]\in\check \HH{}^2(X;\bbZ)\cong \HH^2(X;\bbZ)$

Gauge transformations of the connection do not change the class $[c_{ijk}]\in\check\HH{}^2(X;\mathbb{R})$. To see this consider a gauge transformation which preserves the curvature $F$:
\begin{equation}
A'_i = A_i+\d\epsilon_i, \quad \epsilon_i \in \cC^{\infty}(\cU^i;\bbR).
\end{equation}
This transformation induces a deformation of the $\gamma_{ij}$ as
\begin{equation}\label{eq:cech_gauge}
\gamma'_{ij} = \gamma_{ij}+\epsilon_i-\epsilon_j + \kappa_{ij}
	= \gamma_{ij} + (\dch \epsilon)_{ij} + \kappa_{ij} ,
\end{equation}
where $\kappa_{ij} \in \bbR(\cU^{ij})$ are constants. Thus $c' = \dch \gamma' = \dch \gamma + \dch \kappa = c+\dch \kappa$ is shifted by a \v{C}ech co-boundary, and so $[c'_{ijk}] = [c_{ijk}] \in \check \HH{}^2(X;\bbR)$ is unchanged. Note that in the case of a quantised flux, the integrality of $c_{ijk}$ holds only in a preferred set of gauges for $\gamma_{ij}$ under shifts by real constants $\kappa_{ij}$.

Now consider a general deformation of the above system. We denote the variations by $\Delta (\ldots)$: for example $A' = A + \Delta A$. We fix the (integral) cohomology class of the quantised flux so that $\Delta c_{ijk} = 0$. This means that we must have $(\dch \Delta \gamma)_{ijk} = 0$ and thus as $\cC^{\infty}(\bbR)$ is acyclic we can find $\epsilon_i \in \cC^{\infty}(\cU^i;\bbR)$ with $\Delta\gamma_{ij} = (\dch \epsilon)_{ij}=\epsilon_i - \epsilon_j$. We see the $\gamma_{ij}$ can be deformed only by a gauge transformation, as in \eqref{eq:cech_gauge}. Performing a global gauge transformation $A''_i = A'_i - \d \epsilon_i$, we find that in the new gauge we have
\begin{equation}
\tilde\Delta A_i = A''_i - A_i = \Delta A_i - \d \epsilon_i ,
\end{equation}
so that on $\cU^{ij}$ we have
\begin{equation}
\tilde\Delta A_i - \tilde\Delta A_j = \d (\Delta \gamma_{ij} - \epsilon_i + \epsilon_j) = 0.
\end{equation}
We have shown there exists a gauge in which the variation of the connection is a global one-form $\tilde\Delta A_i = \tilde\Delta A_j$. 

Note that we could have performed the deformation requiring only that $\Delta c_{ijk} = (\dch \kappa)_{ijk}$ for $\kappa_{ij} \in \bbR(\cU^{ij})$ -- this fixes the real cohomology class $[c_{ijk}]\in\check\HH{}^2(X;\mathbb{R})$. We would then have deduced that $\Delta\gamma_{ij} = (\dch \epsilon)_{ij} + \kappa_{ij}$, leading to the same gauge transformation of $A$ as above. This would correspond to deforming away from the gauge (choice of $\gamma_{ij}$) in which $c_{ijk}$ are explicitly integral, while above we restricted ourselves to the gauge with $c_{ijk} \in \bbZ(\cU^{ijk})$. 

The story becomes more intricate for non-abelian bundles. Recall that a non-abelian connection $A$ on overlaps ${\cal U}^{ij}$ transforms as
\begin{equation}
A_j=g_{ij}A_i g^{-1}_{ij}+g_{ij}\d g^{-1}_{ij}.
\end{equation}
For the purpose of the present paper we will assume without proof that we can take $\Delta g_{ij}=0$, as in the abelian case. 

\subsection{The two-form gerbe example}
The case of a two-form gerbe is a direct generalisation of the abelian bundle. The gerbe is specified by a set of two-forms $B_i\in\Omega^{2}({\cal U}^{i})$ covering the manifold. The field strength $H = \dd B$ is globally defined. As before, this means that on overlaps ${\cal U}^{ij}$ we have
\begin{equation}
\label{eq:gaugeCh}
B_i-B_j=\d\lambda_{ij},
\end{equation}
for some one-forms $\lambda_{ij}\in\Omega^{1}({\cal U}^{ij})$. Again we have $(\dch B)_{ij}=\d\lambda_{ij}$ on ${\cal U}^{ij}$ so that $(\dch\lambda)_{ijk}$ is $\dd$-closed
\begin{equation}
\d(\dch\lambda)_{ijk} =  (\dch^2 B)_{ijk}= 0.
\end{equation}
As ${\cal U}^{ij}$ is contractible $(\dch\lambda)_{ijk}$ is actually exact
\begin{equation}
(\dch\lambda)_{ijk}=\d\zeta_{ijk},
\end{equation}
for some $\zeta_{ijk}\in\cC^{\infty}({\cal U}^{ijk};\bbR)$. Finally we take the co-cycle of $\d\zeta_{ijkl}$ on quadruple overlaps ${\cal U}^{ijkl}$:
\begin{equation}
\label{eq:Const2}
\d(\dch\zeta)_{ijkl} = (\dch^2\lambda)_{ijkl} = 0.
\end{equation}
It follows that the functions $c_{ijkl}=(\dch\zeta)_{ijkl}$ are constants. As before, if the flux is quantised the $c_{ijkl}$ can be made integral by appropriate choice of $\zeta_{ijk}$. Indeed, the $c_{ijkl}$ represent a co-cycle in the sheaf cohomology
\begin{equation}
[c_{ijkl}]\in\check \HH{}^3(X;\mathbb{Z})\cong \HH^3(X;\mathbb{Z}),
\end{equation}
where the class in $\HH^3(X;\mathbb{Z})$ is given by the flux $H=\d B$.

Consider deformations of the above system. From the quantisation of the constants $c_{ijkl}$ and the acyclicity of $\cC^{\infty}(\bbR)$ we have
\begin{equation}
\label{eq:varConst2}
\Delta c_{ijkl} = (\dch\Delta\zeta)_{ijk} = 0 
\quad \Ra \quad
\Delta\zeta_{ijk} = (\dch \kappa)_{ijk},
\end{equation}
for some $\kappa_{ij} \in \cC^{\infty}(\cU^{ij};\bbR)$. This leads immediately to\footnote{Note that again this relation would be unchanged if we simply fixed the real cohomology class $[c]\in\check \HH{}^3(X;\bbR)$ so that $\Delta c_{ijkl} = (\dch r)_{ijkl}$ for some $r_{ijk} \in \bbR(\cU^{ijk})$. We would then have $\Delta\zeta_{ijk} = r_{ijk} + (\dch \kappa)_{ijk}$.}
\begin{equation}
(\dch \Delta \lambda)_{ijk} = \d \Delta\zeta_{ijk} = (\dch \d\kappa)_{ijk}
\quad \Ra \quad
\Delta \lambda_{ij} = \d\kappa_{ij} + (\dch\epsilon)_{ij},
\end{equation}
for some $\epsilon_i \in \Omega^1(\cU^i)$. Putting this all together we have
\begin{equation}
\Delta B_i - \Delta B_j = \d\Delta\lambda_{ij} = \d\epsilon_i - \d\epsilon_j,
\end{equation}
so that
\begin{equation}
\Delta B_i - \d\epsilon_i = \Delta B_j - \d\epsilon_j .
\end{equation}
Looking at these formulae, we see that the variations are forced to take precisely the form needed to constitute a gauge transformation of $B$ and its descendants $\lambda$ and $\zeta$. Thus, in an appropriately chosen gauge, we have that $\Delta B$ is a global two-form.

\subsection{Heterotic flux quantisation}
\label{app:Het-flux}

We now come to the example relevant for the present paper: the $B$ field of the heterotic string. Recall that the anomaly cancellation condition reads
\begin{equation}
\label{eq:defH}
H=\d B+\omega_\text{CS}(A),
\end{equation}
where $\omega_\text{CS}(A)$ denotes the Chern--Simons three-form and we set $\frac{\a}{4}=1$. A gauge transformation of $A$ of the form
\begin{equation}
A' = g A g^{-1}+g\d g^{-1}
\end{equation}
induces the following transformation of the Chern--Simons form~\cite{CMTW14}
\begin{equation}
\omega_\text{CS}(A')  = \omega_\text{CS}(A)+\d\tr(g^{-1}\d g\wedge A)-\d\mu(g),
\end{equation}
where 
\begin{equation}
\d\mu(g)=\tr(g\,\d g^{-1}g\,\d g^{-1}g\,\d g^{-1}).
\end{equation}
Hence we have
\begin{equation}
\omega_\text{CS}(A') = \omega_\text{CS}(A)+\d\omega^2(g,A),
\end{equation}
where
\begin{equation}
\omega^2(g,A)=\tr(g^{-1}\d g\wedge A)-\mu(g).
\end{equation}

Let us now consider the patching on ${\cal U}^{ij}$. From equation \eqref{eq:defH} we get
\begin{equation}
\d(B_i-B_j)+\d\omega^2_{ij}(g,A)=0,
\end{equation}
where
\begin{equation}
\omega^2_{ij}(g,A)=\tr\bigl(g_{ji}^{-1}\d g_{ji}\wedge A_j\bigr)-\mu(g_{ji}).
\end{equation}
It follows that
\begin{equation}
\label{eq:Hetco-cycle1}
B_i-B_j+\omega^2_{ij}(g,A)=\d\lambda_{ij},
\end{equation}
for some one-forms $\lambda_{ij}$. Taking a co-cycle of this on triple intersections ${\cal U}^{ijk}$ gives the relation
\begin{equation}
\label{eq:omega2-co-cycle}
(\dch \omega^2)_{ijk}=\d(\dch\lambda)_{ijk}.
\end{equation}
It can further be shown that 
\begin{equation}
\label{eq:omega2exact}
(\dch \omega^2)_{ijk}(g,A)=\d\omega^1_{ijk}(g),
\end{equation}
for some one-forms $\omega^1_{ijk}(g) \in \Omega^1(\cU^{ijk})$.

The expression for $\omega^1_{ijk}(g)$ is not relevant in the present context, but it can be taken to be independent of the gauge connection $A$, depending on the transition functions $g_{ij}$. To see this, simply vary $\omega^2_{ij}(g,A)$ with respect to $A$. We find that
\begin{equation}
\Delta\omega^2_{ij}(g,A)=\tr(\alpha_i\wedge A_i)-\tr(\alpha_j\wedge A_j),
\end{equation}
where we have defined $\Delta A=\alpha$, which transforms appropriately in the adjoint representation when the transition functions $g_{ij}$ are kept constant. From this it is clear that
\begin{equation}
\Delta(\dch \omega^2)_{ijk}(g,A)=0,
\end{equation}
which shows that we may take $\omega^1_{ijk}(g)$ to be independent of $A$ without loss of generality.

Putting together~\eqref{eq:omega2-co-cycle} and~\eqref{eq:omega2exact}, we then get the relation
\begin{equation}
\label{eq:HetCocycl3}
(\dch\lambda)_{ijk}-\omega^1_{ijk}(g)=\d\zeta_{ijk},
\end{equation}
for some functions $\zeta_{ijk} \in \cC^{\infty}(\cU^{ijk};\bbR)$. We again co-cycle this relation on quadruple intersections ${\cal U}^{ijkl}$ to get
\begin{equation}
\label{eq:trivPont}
\d\bigl(\omega^0_{ijkl}(g)+(\dch\zeta)_{ijkl}\bigr)=0,
\end{equation}
where we have used the fact that
\begin{equation}
(\dch\omega^1)_{ijkl}(g)=\d\omega^0_{ijkl}(g),
\end{equation}
for some functions $\omega^0_{ijkl}(g)$. We can co-cycle this relation again on quintuple overlaps ${\cal U}^{ijklm}$ to get
\begin{equation}
\d(\dch\omega^0)_{ijklm}(g)=0,
\end{equation} 
and thus the numbers $k_{ijklm} = (\dch\omega^0)_{ijklm}(g)$ define a class 
\begin{equation}
[k_{ijklm}]\in\check \HH{}^4(X;\bbR)\cong \HH^4(X;\bbR).
\end{equation}
This represents the Pontryagin class $\tr F\wedge F$ in the cohomology of the sheaf $\bbR$. 
Note however from \eqref{eq:trivPont} we have for our particular bundle
\begin{equation}
\label{eq:defClassC}
\omega^0_{ijkl}(g)+(\dch\zeta)_{ijkl}=c_{ijkl},
\end{equation}
for some constants $c_{ijkl}$. Computing the co-cycle $(\dch\omega^0)_{ijklm}(g)$ we thus find
\begin{equation}
k_{ijklm}=(\dch c)_{ijklm},
\end{equation}
and so $[k_{ijklm}]=0\in\check{\HH}{}^4(X;\bbR)$. This is the sheaf cohomology version of the statement that the bundle in question has a trivial first Pontryagin class as $\tr F\wedge F = \d H$ is exact.

\subsection{Deforming the system and a well-defined global two-form}
\label{app:hetGlobal}
We now consider variations of the above heterotic story. The goal is to show that the heterotic quantisation condition leads us naturally to a global two-form 
\begin{equation}
{\cal B}=\Delta B+\tr(\alpha\wedge A).
\end{equation}
This is an essential part of the moduli of the main text, where $\alpha = \Delta A$ is the global one-form variation of the gauge connection. The story is very similar to that of the two gerbe, with some subtleties. Note that as for the abelian bundle, we will assume that we can choose to keep the transition functions $g_{ij}$ constant under deformations, that is
\begin{equation}
\label{eq:no-vary-g}
\Delta g_{ij}=0,
\end{equation}
even in the case where the bundle is non-abelian. With this assumption, any deformation of the bundle connection $\alpha=\Delta A$ can be assumed to be a section of $\Omega^1(\End V)$.

We begin by noting that, imposing~\eqref{eq:no-vary-g}, a deformation of \eqref{eq:defClassC} gives
\begin{equation}
(\dch \Delta\zeta)_{ijkl} =\Delta c_{ijkl}.
\end{equation}
Hence the constants $\Delta c_{ijkl}$ define a sheaf cohomology class in $\check{\HH}{}^3(X;\bbR)$, even though the original $c_{ijkl}$ did not. 
We will require that this class vanishes $[\Delta c_{ijkl}] = 0 \in \check{\HH}{}^3(X;\bbR)$, so that the deformation does not produce any new third cohomology. One could think of this as fixing the topological data associated to the original numbers $c_{ijkl}$, assuming that they have some notion of ``integrality" which cannot be continuously deformed, even though they do not explicitly define such a class themselves. Indeed, this condition also seems to agree with the general world-sheet arguments on the heterotic $B$ field and flux quantisation made in \cite{Witten00}, where one thinks of the $B$ field as a torsor. The lack of the notion of zero in the space of $B$ fields is reflected in our setup by the lack of an explicit cohomology class associated to the $c_{ijkl}$. However, the variations of $B$ about a given starting point do define a cohomology class naturally.

From this requirement we have
\begin{equation}
(\dch \Delta\zeta)_{ijkl} =(\dch r)_{ijkl}
\quad \Ra \quad 
\Delta\zeta_{ijk} = r_{ijk} + (\dch \kappa)_{ijk},
\end{equation}
for some $r_{ijk} \in \bbR(\cU^{ijk})$ and $\kappa_{ij} \in \cC^{\infty}(\cU^{ij})$. Next we take the variation of~\eqref{eq:HetCocycl3}, again imposing~\eqref{eq:no-vary-g}, and use the acyclicity of $\Omega^1(\cU^{ijk})$ to find
\begin{equation}
(\dch \Delta \lambda)_{ijk} = \d \Delta\zeta_{ijk} = (\dch \d\kappa)_{ijk}
\quad \Ra \quad
\Delta \lambda_{ij} = \d\kappa_{ij} + (\dch\epsilon)_{ij} ,
\end{equation}
exactly as for the simple two gerbe case. Finally, we take the variation of~\eqref{eq:Hetco-cycle1} to obtain
\begin{equation}
\Delta B_i - \d\epsilon_i +\tr(\alpha_i\wedge A_i) 
	= \Delta B_j - \d\epsilon_j +\tr(\alpha_j\wedge A_j).
\end{equation}
We see we can absorb the $\d\epsilon$ terms via a global gauge transformation of $B$ as before, so that on the overlaps ${\cal U}^{ij}$ we have
\begin{equation}
{\cal B}_i={\cal B}_j.
\end{equation}
Thus $\cal B$ defines a global two-form. Note in particular that $\cal B$ is gauge invariant with respect to gauge transformations of the bundle~\cite{COM17}. This follows from the Green--Schwarz mechanism wherein the $B$ field transforms as
\begin{equation}
B\rightarrow B-\omega^2(g,A).
\end{equation}

%%%%%%%%%%%%%%%%%%%%%%%%%%%%%%%%%%%%%%%%%%%%%%%%%%%%%%%%%%%%%%%%%%%%%%%%%%

\section{The off-shell \texorpdfstring{$\mathcal{N}=1$}{N = 1} parameter space and holomorphicity of \texorpdfstring{$\Omega$}{Omega}}
\label{app:N=1}

In this appendix, we present a description of the space of off-shell scalar field configurations that appears when we rewrite the ten-dimensional theory with manifest four-dimensional $\mathcal{N}=1$ supersymmetry. This rewriting is done in the same spirit as the rewriting of eleven-dimensional supergravity as a four-dimensional $\mathcal{N}=8$ theory in~\cite{Nicolai87} and the rewriting of type II supergravities as four-dimensional $\mathcal{N}=2$ theories in~\cite{GLW06,GLW07,GLSW09,AW15,AW15b}. Here we focus only on the scalars, which are described as an (infinite-dimensional) space of generalised geometric structures of a type we will specify. We then show how to recover the three-form $\Omega$ from the generalised geometric structure and outline how one can see that it is holomorphic on the off-shell field space, as we claim in section~\ref{sec:par_defs}.

%%%%%%%%%%%%%%%%%%%%%%%%%

\subsection{\texorpdfstring{$\SU{3}\times\SO{6}$}{SU(3) x SO(6)} structures in the NS-NS sector}
\label{app:O66}

As a warm up, we consider the common NS-NS sector of ten-dimensional string theories, which is well known to admit a description in the language of $\SO{10,10}\times\bbR^+$ generalised geometry~\cite{CSW11b} (see also~\cite{Siegel93a,HHZ10a}). When rewriting the theory as a four-dimensional theory, we consider the spacetime to admit a product structure, breaking the ten-dimensional Lorentz symmetry to $\SO{3,1}\times\SO{6}$. The fields which are naturally $\SO{3,1}$ scalars can be packaged as the generalised metric of an $\SO{6,6}\times\bbR^+$ generalised geometry on the six internal spatial dimensions (see for example~\cite{GLW06,GLW07}). 

As discussed in~\cite{CSW14b}, the conditions for an $\mathcal{N}=1$ supersymmetric vacuum can be phrased as the existence of an integrable $\SU{3}\times\SO{6}$ structure on the internal six-dimensional generalised tangent space $E\simeq TX \oplus T^*X$. Such a structure encodes the generalised metric and thus the physical fields. If we require only the presence of an off-shell $\mathcal{N}=1$ supersymmetry, then one merely restricts the fields to admit a globally defined spinor, corresponding to a (possibly non-integrable) $\SU{3}\times\SO{6}$ structure. The configuration space of off-shell scalar fields that we require in our four-dimensional $\mathcal{N}=1$ description of the ten-dimensional theory is thus the (infinite-dimensional) space of $\SU{3}\times\SO{6}$ structures on the generalised tangent space $E$. This is the $\mathcal{N}=1$ NS-NS sector analogue of the discussion of the $\mathcal{N}=2$ vector- and hyper-multiplet structures given for the full type II theories in~\cite{GLSW09,AW15}. 

An $\SU{3}\times\SO{6} \subset \SO{6,6}\times\bbR^+$ structure on $E$ is specified by a particular type of complex section $\gtft$ of 
\begin{equation}
	W = L \otimes \ext^3 E_\bbC ,
\end{equation}
where $L\simeq \ext^6 T^*X$ is the auxiliary $\bbR^+$ bundle transforming in the $\rep{1_{+1}}$ representation of $\SO{6,6}\times\bbR^+$ as introduced in~\cite{CSW11b}. It is convenient to write $\gtft$ as 
\begin{equation}
	\gtft = \Phi \gtf, \qquad
	\gtf \in \Gamma(\ext^3 E_\bbC),
\end{equation}
where $\Phi \in \Gamma(L)$ is the generalised density with $\Phi = \sqrt{g}\ee^{-2\phi}$ in the coordinate frame (see~\cite{CSW11b}). In order for $\gtft$ to have the correct stabiliser, $\gtf$ must lie in a particular orbit under the action of $\SO{6,6}$ on $\ext^3 E_\bbC$ at each point of the manifold $X$.

One can alternatively describe such a generalised structure in terms of a generalised metric, which defines an $\SO{6}\times\SO{6}$ structure on $E$, which then splits as $E\simeq C_+ \oplus C_-$, so that
\begin{equation}
	\ext^3E \longrightarrow 
		\ext^3C_+ \oplus (\ext^2C_+\otimes C_-) 
			\oplus (C_+\otimes\ext^2C_-) \oplus \ext^3C_-
\end{equation}
To this, one must add a non-vanishing section of the spin bundle for $C_+$ which we denote $\eta \in \Gamma(S(C_+))$. Using local bases $\hE^+_m$ for $C_+$ and $\hE^-_{\tm}$ for $C_-$ defined by\footnote{These are the split frames of~\cite{CSW11b}, but constructed from an arbitrary local frame for the tangent bundle rather than a vielbein for the ordinary metric. However, the ordinary metric $g$ is still used to lower the indices on the one-form frames $e^{\pm m}$, such that the $O(6,6)$ metric has components
\begin{equation}
	\eta = \begin{pmatrix} g & 0 \\ 0 & -g \end{pmatrix}
\end{equation}
in these frames.
}
\begin{equation}
\begin{aligned}
	\hE^+_m &= \hat{e}^+_m + e^+_m - i_{\hat{e}^+_m} B , \\
	\hE^+_{\tm} &= \hat{e}^-_{\tm} - e^-_{\tm} - i_{\hat{e}^-_{\tm}} B ,
\end{aligned}
\end{equation}
one can write an explicit formula for the object $\gtf$
\begin{equation}
	\gtf = \tfrac{1}{3!} (\eta^\text{T} \gamma^{mnp} \eta) 
		\hE^+_m \wedge \hE^+_n \wedge \hE^+_p
		= \tfrac{1}{3!}\, \Psi^{mnp} \hE^+_m \wedge \hE^+_n \wedge \hE^+_p ,
\end{equation}
where $\Psi$ is the three-form spinor bilinear of $\eta$ with itself. Writing the object $\gtf$ in this way, it is guaranteed that it will lie in the correct orbit and its stabiliser is $\SU{3}\times\SO{6}$.

Given an $\mathcal{N}=2$ structure parametrised by two complex pure spinors $\Phi^\pm \in S(E) \otimes L^{1/2}$ (see \cite{GLW06,GMPT04b}), one can also build an $\mathcal{N}=1$ structure via the expression
\begin{equation}
	\gtft^{MNP} = \bar\Phi^+ \Gamma^{MNP} \Phi^-,
\end{equation}
thus providing a third description of the structure.

Having defined the structure, we will now show how to extract the ordinary complex three-form on the manifold from it. Recall that the generalised tangent space is an extension
\begin{equation}
	0 \longrightarrow T^*X \longrightarrow E \stackrel{\pi}{\longrightarrow} TX \longrightarrow 0,
\end{equation}
with the classes of such extensions labelled by the cohomology class of the three-form flux $[H] \in \HH^3(M;\bbR)$. The map $\pi$ is referred to as the anchor map of the Courant algebroid $E$. This anchor map induces further maps on tensor products of $E$. In particular we obtain an induced map, which we also label $\pi$,
\begin{equation}
	\pi \colon \ext^3 E  \longrightarrow \ext^3 TX.
\end{equation}
Acting on the bundle $L \otimes \ext^3 E$ and using $L \simeq \det T^*X$ we have
\begin{equation}
	\pi \colon L\otimes \ext^3 E  \rightarrow \det T^*X \otimes \ext^3 TX
		\cong \ext^3 T^*X.
\end{equation}
Thus, applying the anchor map to the generalised structure $\gtft$ we obtain an ordinary three-form on the manifold. 

We now calculate this three-form explicitly. We first note that
\begin{equation}
	\pi (\hE^\pm_m) = \hat{e}^\pm_m ,
\end{equation}
which immediately gives us
\begin{equation}
	\pi(\gtf) 
		= \tfrac{1}{3!}\, \Psi^{mnp} (\hat{e}_m \wedge \hat{e}_n \wedge \hat{e}_p)
			\in \ext^3 TX_\bbC
\end{equation}
However, we are interested in the object $\gtft = \Phi \gtf$ and in our frame we have $\Phi = \sqrt{g} \ee^{-2\phi}$ so that
\begin{equation}
	\pi(\gtft) 
		= \tfrac{1}{3!} \sqrt{g} \ee^{-2\phi} \Psi^{mnp} 
			(\hat{e}_m \wedge \hat{e}_n \wedge \hat{e}_p) 
				\in \ext^6T^*X \otimes \ext^3 TX_\bbC
\end{equation}
Now we use the natural isomorphism
\begin{equation}
\label{eq:density-isom}
\begin{aligned}
	\ext^6T^*X \otimes \ext^3 TX &\longrightarrow \ext^3 T^*X \\
	\tfrac{1}{3!} X^{mnp} (\hat{e}_m \wedge \hat{e}_n \wedge \hat{e}_p) 
		&\longmapsto
			\tfrac{1}{(3!)^2} \epsilon_{mnpm'n'p'} X^{m'n'p'} (e^m\wedge e^n \wedge e^p)
\end{aligned}
\end{equation}
where $\epsilon_{m_1 \dots m_6} (=\pm1)$ is the Levi-Civita symbol. The standard metric volume form is then $\varepsilon_{m_1 \dots m_6} = \sqrt{g} \epsilon_{m_1 \dots m_6}$ and we find that under the identification~\eqref{eq:density-isom} we have
\begin{equation}
	\pi(\gtft) 
		=  \ee^{-2\phi} \tfrac{1}{(3!)^2} \varepsilon_{mnpm'n'p'}  \Psi^{m'n'p'} 
			(e^m\wedge e^n \wedge e^p)
		= \ee^{-2\phi} (\star \Psi)
		= -\ii\,  \ee^{-2\phi} \Psi.
\end{equation}

To conclude our discussion of the $\SU{3}\times\SO{6}$ structure in $\SO{6,6}\times\bbR^+$ generalised geometry, we note that standard group theoretical arguments give us that the homogeneous space
\begin{equation}
\label{eq:N=1coset}
	\frac{\SO{6,6}\times\bbR^+}{\SU{3}\times\SO{6}}
\end{equation}
is diffeomorphic to the orbit of $\gtft$ (at a point in $M$) under the action of $\SO{6,6}\times\bbR^+$ on the complex $\rep{220}_\bbC$ representation. The homogeneous space~\eqref{eq:N=1coset} has a complex structure, with respect to which the element $\gtft$ of the $\rep{220}_\bbC$ is holomorphic, that is the embedding of~\eqref{eq:N=1coset} into $\bbC^{220}$ is a holomorphic map. If we imagine that this complex structure naturally extends to the infinite-dimensional space of $\SU{3}\times\SO{6}$ structures on $E$, then we expect that our generalised three-form $\gtft$ will be holomorphic on that space. As the anchor map $\pi$ is linear and fixed by the topology of $E$, we have that $\pi(\gtft) =- \ii\,\ee^{-2\phi}\Psi =- \ii\,\Omega$ is also holomorphic on the space of such structures.

Note that the decomposition of the Lie algebras appearing here is
\begin{equation}
	\mathfrak{so}(6,6) \ra 
		\mathfrak{su}(3) \oplus \mathfrak{u}(1) \oplus \mathfrak{so}(6)
		\oplus \Big[ \repp{\bar3}{1}_{+2} \oplus \repp{3}{1}_{-2} \Big]_\bbR 
		\oplus \Big[ \repp{3}{6}_{+1}  \oplus \repp{\bar3}{6}_{-1} \Big]_\bbR,
\end{equation}
so that we can locally parametrise the space~\eqref{eq:N=1coset} by exponentiating the action of the complex Lie algebra elements which do not annihilate $\gtft$. Taking out the overall scale we have
\begin{equation}
	\gtft' = \ee^c \ee^{\alpha+\beta} \cdot \gtft,
	\qquad
	c \in \bbC^* 
	\quad \alpha \in \repp{\bar3}{1}_{+2}
	\quad \beta \in \repp{3}{6}_{+1} .
\end{equation}
The parameters $c, \alpha$ and $\beta$ then become local complex coordinates on the space~\eqref{eq:N=1coset}. Note that as $\SU{3}$ objects, $\alpha$ and $\beta$ carry the same degrees of freedom as the parameters $(\mu, (\Delta \omega + \ii\,\Delta  B)_{(1,1)}, (\Delta \omega + \ii\,\Delta B)_{(0,2)})$ used to parametrise the space of $\mathcal{N}=1$ field configurations in sections~\ref{sec:par_defs} and~\ref{sec:constraints-SU3-holomorphicity} of the main text, while the parameter $c$ is associated to K\"ahler transformations.

%%%%%%%%%%%%%%%%%%%%%%%%%

\subsection{\texorpdfstring{$\SU{3}\times\SO{6+n}$}{SU(3) x SO(6+n)} structures in heterotic supergravity}
\label{app:O66+gauge}

We can perform a similar analysis to the above in $\SO{6,6+n}\times\bbR^+$ generalised geometry for heterotic supergravity~\cite{Garcia-Fernandez:2013gja,CMTW14}, where $n$ is the dimension of the gauge group $G$. In that geometry the generalised tangent space $E$ is an extension of the real Atiyah algebroid $A$
\begin{equation}
	0 \longrightarrow \mathcal{C}^\infty (\mathfrak{g}) \longrightarrow A 
		\longrightarrow TX \longrightarrow 0,
\end{equation}
by the cotangent bundle
\begin{equation}
	0 \longrightarrow T^*X \longrightarrow E \longrightarrow A \longrightarrow 0,
\end{equation}
with the composition of the maps above still giving an anchor map $\pi \colon E \ra TX$.

The structure group of the bundle $E$ is then the geometric subgroup $(\GL{6,\bbR}\times G) \ltimes ((T^*X\otimes\mathfrak{g}) \ltimes \ext^2 T^*X)$ of $\SO{6,6+n}\times\bbR^+$, though as usual in generalised geometry we think of $E$ as an $\SO{6,6+n}\times\bbR^+$ vector bundle. The generalised metric defines an $\SO{6}\times\SO{6+n}$ structure on $E$, with a local frame $(\hE^+_m, \hE^-_{\tm}, \hE^-_\alpha)$ which can be built from the physical fields $(g,B,\phi,A)$ (see~\cite{CMTW14} for details). 
The presence of a single globally defined spinor on $M$ breaks the $\SO{6}$ factor to $\SU{3}$, so that an $\mathcal{N}=1$ structure is an $\SU{3}\times\SO{6+n}$ structure on $E$. 
Using the split frame, one can again write an explicit formula for an object $\gtft \in L \otimes\ext^3 E$ defining such an $\mathcal{N}=1$ structure:
\begin{equation}
	\gtft = \tfrac{1}{3!} \Phi (\eta^\text{T} \gamma^{mnp} \eta) 
		\hE^+_m \wedge \hE^+_n \wedge \hE^+_p
		= \tfrac{1}{3!} \sqrt{g} \ee^{-2\phi} \Psi^{mnp} \hE^+_m \wedge \hE^+_n \wedge \hE^+_p .
\end{equation}
As we still have $\pi (\hE^+_m) = \hat{e}^+_m$, the argument of appendix~\ref{app:O66} still holds to give us that $ \pi(\gtft) = - \ii\,  \ee^{-2\phi} \Psi$, and similar group theoretical reasoning leads us to conclude that both $\gtft$ and $\ee^{-2\phi} \Psi = \Omega$ are holomorphic on the coset space
\begin{equation}
\label{eq:N=1coset-gauge}
	\frac{\SO{6,6+n}\times\bbR^+}{\SU{3}\times\SO{6+n}}
\end{equation}
at each point of $X$. In this case, the corresponding Lie algebra decomposition reads
\begin{equation}
\begin{split}
	\mathfrak{so}(6,6+n) &\ra 
		\mathfrak{su}(3) \oplus \mathfrak{u}(1) \oplus \mathfrak{so}(6+n)
		\oplus \Big[ \repp{\bar3}{1}_{+2} \oplus \repp{3}{1}_{-2} \Big]_\bbR \\
		& \eqspace \oplus \Big[ \repp{3}{6+n}_{+1}  \oplus \repp{\bar3}{6+n}_{-1} \Big]_\bbR 
\end{split}
\end{equation}
so that we can locally parametrise the orbit of $\gtft$ via
\begin{equation}
	\gtft' = \ee^c \ee^{\alpha+\beta} \cdot \gtft
	\qquad
	c \in \bbC^* 
	\quad \alpha \in \repp{\bar3}{1}_{+2}
	\quad \beta \in \repp{3}{6+n}_{+1} .
\end{equation}
As $\SU{3}\times G$ objects, we have the same degrees of freedom as in appendix~\ref{app:O66}, but now augmented by $\beta_{\ba}{}^\alpha$, that is the $(0,1)$-form part of the deformation of the gauge field. These then match the parametrisation of the off-shell $\mathcal{N}=1$ field space of section~\ref{sec:par_defs} as employed in section~\ref{sec:incl-bundle}.

%%%%%%%%%%%%%%%%%%%%%%%%%%%%%%%%%

\subsection{The off-shell hermitian structure on \texorpdfstring{$V$}{V}}
\label{sec:hermitian}

The off-shell parameter space of the Hull–Strominger system is the space of $\SU3$ structures, $B$ fields and gauge fields for $G$. On-shell, we know that the gauge bundle $V$ must be a polystable holomorphic bundle. We outline which parts of this structure on the gauge bundle survive off-shell. 

The gauge group $G$ is a compact unitary \emph{real} Lie group, which has complex representations. 
The vector bundle $V_\mathbb{C}$ is a \emph{complex} vector bundle with structure group $G_{\mathbb{C}}$, the complexification of $G$. 
On-shell (on a solution of the Hull-Strominger system), this complex vector bundle has two structures on it: a holomorphic structure and a hermitian structure.

The hermitian structure on $V_\mathbb{C}$ is a reduction of the structure group $G_{\mathbb{C}}$ to the compact real form $G$. This is given by a set of local $G_{\mathbb{C}}$ frames for $V_\mathbb{C}$ on patches of $X$, which, on the overlaps of patches, are related by transition functions in $G \subset G_\bbC$ only. The holomorphic structure on $V_\mathbb{C}$ is given by a set of $G_{\mathbb{C}}$ frames with respect to which the transition functions are holomorphic functions into the complexified group $G_\bbC$. Clearly these two sets of frames are different for the simple reason that there are no holomorphic maps into the real group $G$.

From the complex vector bundle $V_\mathbb{C}$, we can define a real vector bundle
\begin{equation}
V=[V_\mathbb{C}\oplus\bar{V}_\mathbb{C}]_{\mathbb{R}},
\end{equation}
which also admits an action of the complex group $G_\bbC$. The hermitian structure is a positive definite metric on $V$ which pairs $V_\mathbb{C}$ and $\bar{V}_\mathbb{C}$, such that in the special frames alluded to above its component matrix takes the canonical form
\begin{equation}
	h = \frac12 \begin{pmatrix} 0 & \delta_{\alpha \bar\beta} \\ 
		\delta_{\bar\alpha \beta} & 0 \end{pmatrix}
\end{equation}
where $\alpha,\beta = 1,\ldots , \dim_\bbC V_\mathbb{C}$ are indices for $V_\mathbb{C}$ and $\bar\alpha,\bar\beta = 1,\ldots , \dim_\bbC V_\mathbb{C}$ are indices for $\bar{V}_\mathbb{C}$.

Off-shell, the manifold $X$ admits an almost complex structure as part of the $\SU3$ structure. As $X$ is not honestly complex, we lose the holomorphic structure on $V_\mathbb{C}$ – we cannot define holomorphic maps without an integrable complex structure. Nevertheless, physics tells us we have a real connection on $V_\bbR$ and that the physical gauge group $G$ is compact and unitary. This means we have the hermitian structure, even off-shell.\footnote{As $X$ is only almost complex, one might be tempted to call this an almost hermitian structure to emphasise that the holomorphic structure is not present.} This hermitian structure simply says that $V_\mathbb{C}$ defines a real vector bundle $V$ with a compact unitary structure group.

The physical connection $A$ is a local section of
\begin{equation}
\Omega^{1}(X;\End V)\sim\Omega^{1}(X;\mathfrak{g}).
\end{equation}
The almost complex structure defines a split of this into $(1,0)$- and $(0,1)$-form parts. 
As the $(1,0)$- and $(0,1)$-form summands are intrinsically complex, it is natural to see them as living in the complexified Lie algebra, i.e.~$\Omega^{(1,0)}(X;\mathfrak{g}_\bbC)$ and $\Omega^{(0,1)}(X;\mathfrak{g}_\bbC)$, as these are defined via tensor products over the field $\bbC$. 
Naively, the connection has four parts with indices as
\begin{equation}
\begin{matrix}
(A_{a})^{\alpha}{}_{\beta}
\quad & \quad
(A_{a})^{\bar{\alpha}}{}_{\bar{\beta}} \\
(A_{\bar{a}})^{\alpha}{}_{\beta}
\quad & \quad
(A_{\bar{a}})^{\bar{\alpha}}{}_{\bar{\beta}} \end{matrix}
\end{equation}
However, as the connection is real hermitian, any one of them defines the rest via complex conjugation and multiplication by the hermitian metric on $V_\mathbb{C}$. Explicitly, the real condition fixes
\begin{equation}
\big[ (A_{a})^{\alpha}{}_{\beta} \big]^* = (A_{\bar{a}})^{\bar{\alpha}}{}_{\bar{\beta}} ,
\hs{30pt}
\big[ (A_{a})^{\bar{\alpha}}{}_{\bar{\beta}} \big]^* = (A_{\bar{a}})^{\alpha}{}_{\beta} ,
\end{equation}
while the hermitian condition fixes
\begin{equation}
\big[ (A_{a})^{\alpha}{}_{\beta} \big]^* 
	= - h_{\bar{\beta}\beta}  (A_{\bar{a}})^{\beta}{}_{\alpha} h^{\alpha \bar{\alpha}} , 
\hs{30pt}
\big[ (A_{\bar{a}})^{\bar{\alpha}}{}_{\bar{\beta}} \big]^* 
	= - h_{\beta\bar{\beta}}  (A_{a})^{\bar\beta}{}_{\bar\alpha} h^{\bar{\alpha} \alpha} . 
\end{equation}
Together, these allow us to determine all parts of the connection given only $(A_{\bar{a}})^{\alpha}{}_{\beta}$; for example we have
\begin{equation}
(A_{\bar{a}})^{\bar{\alpha}}{}_{\bar{\beta}}
	= - h_{\bar{\beta}\beta}  (A_{\bar{a}})^{\beta}{}_{\alpha} h^{\alpha \bar{\alpha}} .
\end{equation}

On-shell, the holomorphic structure means that, in each patch, one can choose a $G_\bbC$ gauge where $(A_{\bar{a}})^{\bar{\alpha}}{}_{\bar{\beta}}=0$. Off-shell we cannot do this in general, but we can use the previous identities to write any formula purely in terms of $(A_{\bar{a}})^{\alpha}{}_{\beta}$. 
For example, the terms appearing in the superpotential involve traces, which simplify using identities like
\begin{equation}
	(A_{\bar{a}})^{\beta}{}_{\alpha} (A_{\bar{b}})^{\alpha}{}_{\beta} 
		= (A_{\bar{b}})^{\bar\beta}{}_{\bar\alpha} (A_{\bar{a}})^{\bar\alpha}{}_{\bar\beta} .
\end{equation}
These enable us to write all of the needed expressions using only the objects with indices for $V_\mathbb{C}$ (eliminating the appearance of objects with indices for $\bar{V}_\mathbb{C}$); for example
\begin{equation}
	\omega_{\text{CS}} (A) = \tr (A \wedge \dd A + \tfrac23 A\wedge A\wedge A)
	= 2 \Big[ A^{\alpha}{}_{\beta} \wedge \dd A^{\alpha}{}_{\beta} 
		+ \tfrac23 A^{\alpha}{}_{\beta}\wedge A^{\beta}{}_{\gamma}\wedge 
			A^{\gamma}{}_{\alpha} \Big] .
\end{equation}
so that $\omega_{\text{CS}} (A) \wedge \Omega$ features only the components $(A_{\bar{a}})^\alpha{}_{\beta}$. Equivalently we could have written this expression purely in terms of $\bar{V}_\mathbb{C}$ indices or some combination of the two -- this freedom reflects the fact that it is $V$ that appears in the heterotic system, so a split into $V_\mathbb{C}$ and $\bar{V}_\mathbb{C}$ is somewhat arbitrary. The important point is that it is the (0,1) part of the gauge field that appears. This mirrors the generalised geometry argument from the previous section.

Using the freedom to write the Chern--Simons form using only $V_\mathbb{C}$ indices, one can take the bundle appearing in the deformation complex to be the holomorphic bundle
\begin{equation}\label{eq:holomorphic_bundle}
T^{*(1,0)}X\oplus\End V_\mathbb{C}\oplus T^{(1,0)}X .
\end{equation}
which also appears in \cite{Garcia-Fernandez:2018emx}.

%%%%%%%%%%%%%%%%%%%%%%%%%%%%%%%%%%%%%%%%%%%%%%%%%%%%%%%%%%%%%%%%%%%%%%%%%%

%%%%%%%%%%%%%%%%%%%%%%%%%%%%%%%%%%%%%%%%%%%%%%%%%%%%%%%%%%%%%%%%%%%%%%%%%%

\section{Comments on \texorpdfstring{$D$}{D}-terms}
\label{app:Dterms}

In addition to satisfying the $F$-term conditions derived from the superpotential, a solution should also satisfy the $D$-term conditions:
\begin{align}
\label{eq:confBal}
\d\bigl(e^{-2\phi}\omega\wedge\omega\bigr)&=0,\\
\omega\wedge\omega\wedge F&=0.
\end{align}
The first condition is referred to as the {\it conformally balanced condition}, while the second condition is the Yang--Mills condition. In this appendix we want to show that these conditions impose no extra constraints on the heterotic moduli, given some mild assumptions on the geometry and bundle. This is of course common knowledge from the supergravity point of view~\cite{Weinberg1998}.\footnote{As we mention in the main text, for an $\mathcal{N}=1$ supersymmetric theory in four dimensions, supersymmetry breaking is controlled completely by the $F$-terms when there are no FI parameters. Given a solution to the $F$-term conditions, one can always make a complex gauge transformation to find a solution to the $D$-term conditions on the same orbit. In the heterotic case, this is equivalent to assuming the bundles $V$ and $TX$ are stable.}

\subsection{Massless deformations}

We begin by considering compactifications without bundles. The $D$-term condition of relevance is the conformally balanced condition \eqref{eq:confBal}. Consider first a \emph{massless }deformation of this condition so that $y_{0}=(x_{0},\mu_{0})$ satisfies $\bar{D}y_{0}=0$. Our plan is to show that the $D$-term conditions can be solved order-by-order using the gauge symmetries of $x_{0}$ so that they do not further constrain the moduli.

First we note that deformations of the hermitian form $\omega$ and the dilaton $\phi$ are linked via the $\SU3$ normalisation condition
\begin{equation}
\tfrac{\ii}{8}\Omega\wedge\bar{\Omega}=\tfrac{1}{6}\ee^{-4\phi}\omega\wedge\omega\wedge\omega.
\end{equation}
Remembering $\Delta\omega_{(0,2)}=0$, $\Delta\omega_{(1,1)}=-\ii x$ and $\Delta\omega_{(0,2)}=\imath_{\mu}\omega-\ii\,\imath_{\mu}x$, a massless holomorphic variation of this condition gives
\begin{equation}
\ee^{4\Delta_{0}\phi}\frac{1}{3!}\omega\wedge\omega\wedge\omega=\frac{1}{3!}\omega\wedge\omega\wedge\omega-\frac{\ii}{2}\omega\wedge\omega\wedge x_{0}-\frac{1}{2}\omega\wedge x_{0}\wedge x_{0}+\frac{\ii}{3!}x_{0}\wedge x_{0}\wedge x_{0}.
\end{equation}
Now we expand the massless deformation in terms of a small parameter $\epsilon$:
\begin{align}
y_{0} & =\epsilon\,y_{(1)}+\epsilon^{2}\,y_{(2)}+\ldots\\
\Delta_{0}\phi & =\epsilon\,\phi_{(1)}+\epsilon^{2}\,\phi_{(2)}+\ldots
\end{align}
with a corresponding expansion for $x$ and $\mu$. At first and second order the $\SU3$ normalisation condition fixes
\begin{align}
\phi_{(1)} & =-\frac{3}{4}\ii\,\omega\lrcorner x_{(1)},\label{eq:dil_first}\\
\phi_{(2)} & =-\frac{3}{4}\ii\,\omega\lrcorner x_{(2)}+\frac{1}{8}x_{(1)}\lrcorner x_{(1)}.\label{eq:dil_second}
\end{align}
This continues to higher order – the deformation of the dilaton at $n^{\text{th}}$ order is fixed by the non-primitive part of $x_{(n)}$ and the lower-order fields $x_{(i)}$ for $i<n$.

Now consider a massless holomorphic deformation of the conformally balanced condition. We expand in $\epsilon$ and separate into complex type. At $\mathcal{O}(\epsilon^{1})$ the deformation of the conformally balanced condition reduces to
\begin{align}
\partial\Bigl(\ii\,\ee^{-2\phi}\omega\wedge x_{(1)}+\phi_{(1)}\ee^{-2\phi}\omega\wedge\omega\Bigr) & =0,\\
\bar{\partial}\Bigl(\ii\,\ee^{-2\phi}\omega\wedge x_{(1)}+\phi_{(1)}\ee^{-2\phi}\omega\wedge\omega\Bigr) & =\partial\Bigl(\ee^{-2\phi}\omega\wedge\imath_{\mu_{(1)}}\omega\Bigr).
\end{align}
Note now that the Gauduchon metric defined by $\tilde{\omega}=\ee^{-\phi}\omega$ is \emph{balanced} as $\tilde{\omega}\wedge\tilde{\omega}$ is $\dd$-closed. We do this as on a hermitian manifold the various Laplace operators for a balanced metric agree on functions~\cite{Gauduchon1984}:
\begin{equation}
\tilde{\Delta}_{\dd}f=2\tilde{\Delta}_{\partial}f=2\tilde{\Delta}_{\bar{\partial}}f.
\end{equation}
We also use that the Hodge stars\footnote{We are using the convention for the Hodge star where $\alpha\wedge\star\beta=\alpha\lrcorner\beta\,\vol$ so that $\star\omega=\tfrac{1}{2}\omega\wedge\omega$ and $\star \Omega = - \ii\,\Omega$. The dual of a primitive $(1,1)$-form $\alpha_{\text{p}}$ satisfying $\omega\lrcorner\alpha_{\text{p}}$ is $\star\alpha_{\text{p}}=-\omega\wedge\alpha_{\text{p}}$. We also have $\star\alpha_{20}=\alpha_{20}\wedge\omega$ where $\alpha_{20}$ is a $(2,0)$-form. This choice satisfies $\star^{2}=(-1)^{p}$ on a $p$-form. The adjoint Dolbeault operators are defined by $\partial^{\dagger}=-\star\bar{\partial}\star$, and we denote the corresponding operators for the Gauduchon metric with a tilde.} on a $p$-form are related by
\begin{equation}
\star=\ee^{(3-p)\phi}\tilde{\star}.
\end{equation}
Using this we can write the previous equations as
\begin{align}
\bar{\partial}^{\tilde{\dagger}}X_{(1)} & =0,\label{eq:confBaldelbar}\\
\partial^{\tilde{\dagger}}X_{(1)} & =\ii\,\bar{\partial}^{\tilde{\dagger}}\imath_{\mu_{(1)}}\tilde{\omega},\label{eq:confBlaldel}
\end{align}
where we have used the relation between the trace of $x_{(1)}$ and $\phi_{(1)}$ given in (\ref{eq:dil_first}), and we have defined
\begin{equation}
X_{(1)}=\ee^{-\phi}(x_{(1)}-\tfrac{1}{2}\tilde{\omega}\lrcorner x_{(1)}\,\tilde{\omega}).
\end{equation}
The Hodge decomposition for Aeppli cohomology implies that (\ref{eq:confBlaldel}) and (\ref{eq:confBaldelbar}) determine the $(\partial+\bar{\partial})$-exact part of $X_{(1)}$. Indeed an equivalent set of equations is
\begin{align}
\partial^{\tilde{\dagger}}\bar{\partial}^{\tilde{\dagger}}X_{1} & =0,\label{eq:first_non_constant}\\
\partial\bar{\partial}^{\tilde{\dagger}}X_{1} & =0,\label{eq:second_non_constant}\\
\bar{\partial}\partial^{\tilde{\dagger}}X_{1} & =\ii\,\bar{\partial}\bar{\partial}^{\tilde{\dagger}}\imath_{\mu_{1}}\tilde{\omega}.\label{eq:third_non_constant}
\end{align}
We now want to argue that these conditions are simply gauge conditions and so do not impose extra conditions on the moduli. Recalling the form of $\bar{D}y_{0}$ from \eqref{eq:Dbar_definition} we see shifts of $x_{0}$ by $\bar{\partial}$-exact terms drop out explicitly and that shifts by $\partial$-exact terms fall out as we are working modulo $\partial$-exact forms.\footnote{One can also do this calculation with $y$ and $b$ so that the shift by $\partial$-exact forms is explicit too.} A gauge choice for $x_{0}$ then amounts to a choice of element in $\partial\Omega^{(0,1)}(X)+\bar{\partial}\Omega^{(1,0)}(X)$. We make a simplification: let us assume the following cohomologies vanish
\begin{equation}
\text{H}_{\bar\partial}^{(1,0)}(X)=\text{H}_{\bar\partial}^{(2,0)}(X)=\text{H}_{\bar\partial}^{(0,1)}(X)=\text{H}_{\bar\partial}^{(0,2)}(X)=0,
\end{equation}
so that we get a Hodge decomposition of the space of $(\partial+\bar{\partial})$-exact forms as
\begin{equation}
\label{eq:exactHodge}
\partial\Omega^{(0,1)}+\bar{\partial}\Omega^{(1,0)}=\partial\bar{\partial}\Omega^{(0,0)}\oplus\bar{\partial}\partial^{\tilde{\dagger}}\Omega^{(2,0)}\oplus\partial\bar{\partial}^{\tilde{\dagger}}\Omega^{(0,2)}.
\end{equation}
Note that it is important that we include the dilaton degrees of freedom: shifts of $x_{(1)}$ are generically not primitive and so they will change the $\SU3$ normalisation condition, but we can compensate for this by shifting $\phi_{(1)}$ (which does not appear explicitly in the first-order conformally balanced condition).

We start by shifting $x_{(1)}$ by $-\partial\bar{\partial}\kappa_{(1)}$, where $\kappa_{(1)}$ is a function. A short calculation shows that equation (\ref{eq:first_non_constant}) becomes
\begin{equation}
\partial^{\tilde{\dagger}}\bar{\partial}^{\tilde{\dagger}}\Bigl(\ee^{-\phi}(x_{(1)}-\tfrac{1}{2}\tilde{\omega}\lrcorner x_{(1)}\,\tilde{\omega})\Bigr)=\partial^{\tilde{\dagger}}\bar{\partial}^{\tilde{\dagger}}\Bigl(\ee^{-\phi}\partial\bar{\partial}\kappa_{(1)}\Bigr)+\tfrac{1}{2}\tilde{\Delta}_{\partial}\Bigl(\ee^{-\phi}\tilde{\Delta}_{\bar{\partial}}\kappa_{(1)}\Bigr).\label{eq:first_non_constant_operator}
\end{equation}
One can check that the operator acting on $\kappa_{(1)}$ is a positive semi-definite self-adjoint elliptic operator whose image is given by non-constant functions. This means that \label{eq:second_non_constant_operator}(\ref{eq:first_non_constant_operator}) can always be solved by an appropriate choice of $\kappa_{(1)}$. This fixes the $\partial\bar{\partial}$-gauge symmetry of $x_{(1)}$.

Next consider a shift $x_{(1)}$ by $-\bar{\partial}\partial^{\tilde{\dagger}}\alpha_{(1)}$,\footnote{One should think of this $x_{(1)}$ as already gauge transformed to solve the previous condition.} where $\alpha_{(1)}$ is a $(2,0)$-form. A short calculation shows that (\ref{eq:second_non_constant}) becomes
\begin{equation}
\partial\bar{\partial}^{\tilde{\dagger}}\Bigl(\ee^{-\phi}(x_{(1)}-\tfrac{1}{2}\tilde{\omega}\lrcorner x_{(1)}\,\tilde{\omega})\Bigr)=\partial\bar{\partial}^{\tilde{\dagger}}(\ee^{-\phi}\bar{\partial}\partial^{\tilde{\dagger}}\alpha_{(1)}).
\end{equation}
Again, one can check that the operator acting on $\alpha_{(1)}$ is positive semi-definite, self-adjoint and elliptic so that (\ref{eq:second_non_constant_operator}) can always be solved for by a choice of $\alpha_{(1)}$. This fixes the $\bar{\partial}\partial^{\tilde{\dagger}}$-gauge symmetry of $x_{(1)}$.

Finally consider a shift of $x_{(1)}$ by $-\partial\bar{\partial}^{\tilde{\dagger}}\beta_{(1)}$, where $\beta_{(1)}$ is a $(0,2)$-form. A short calculation shows that (\ref{eq:third_non_constant}) becomes 
\begin{equation}
\bar{\partial}\partial^{\tilde{\dagger}}\Bigl(\ee^{-\phi}(x_{(1)}-\tfrac{1}{2}\tilde{\omega}\lrcorner x_{(1)}\,\tilde{\omega})\Bigr)-\ii\,\bar{\partial}\bar{\partial}^{\tilde{\dagger}}(\imath_{\mu_{(1)}}\tilde{\omega})=\bar{\partial}\partial^{\tilde{\dagger}}\Bigl(\ee^{-\phi}\partial\bar{\partial}^{\tilde{\dagger}}\beta_{(1)}\Bigr).
\end{equation}
As $\imath_{\mu_{(1)}}\omega$ is a $(0,2)$-form and we assume $\text{H}^{(0,2)}(X)$ vanishes, $\bar{\partial}\bar{\partial}^{\tilde{\dagger}}\imath_{\mu_{(1)}}\omega$ is actually $\bar{\partial}\partial^{\tilde{\dagger}}$-exact. Thus when we shift $x_{(1)}$, this equation can be solved providing the operator acting on the gauge parameter is elliptic and positive semi-definite as before – it is simple to check that this is the case. This fixes the $\partial\bar{\partial}^{\tilde{\dagger}}$-gauge symmetry of $x_{(1)}$.

What happens at higher orders in $\epsilon$? At second order we have
\begin{align}
\bar{\partial}^{\tilde{\dagger}}(X_{(2)}+A_{(2)}) & =0,\label{eq:second_first}\\
\partial^{\tilde{\dagger}}(X_{(2)}+A_{(2)}) & =\ii\,\bar{\partial}^{\tilde{\dagger}}\imath_{\mu_{(2)}}\tilde{\omega}+\bar{\partial}^{\tilde{\dagger}}B_{(2)},\label{eq:second_second}
\end{align}
where
\begin{align}
X_{(2)}(x_{(2)}) & =\ee^{-\phi}(x_{(2)}-\tfrac{1}{2}\tilde{\omega}\lrcorner x_{(2)}\,\tilde{\omega}).\\
A_{(2)}(x_{(1)}) & =\ii\,\ee^{-\phi}\Bigl(-\tfrac{3}{8}(\omega\lrcorner x_{(1)})^{2}\omega+\tfrac{1}{4}x_{(1)}\lrcorner x_{(1)}\,\omega+\tfrac{1}{2}\omega\lrcorner x_{(1)}\,x_{(1)}+\star(x_{(1)}\wedge x_{(1)})\Bigr),\\
B_{(2)}(x_{(1)},\mu_{(1)}) & =\ee^{-\phi}\Bigl(-\tfrac{1}{2}\omega\lrcorner x_{(1)}\,\imath_{\mu_{(1)}}\omega+\imath_{\mu_{(1)}}x_{(1)}+\star(x_{(1)}\wedge\imath_{\mu_{(1)}}\omega)\Bigr).
\end{align}
We see that $X_{(2)}$ depends only on the second-order correction to $x_{0}$ while $A_{(2)}$ and $B_{(2)}$ are fixed by the first-order terms (which should be thought of as gauge transformed to solve the first-order conditions). Again (\ref{eq:second_first}) and (\ref{eq:second_second}) are equivalent to 
\begin{align}
\partial^{\tilde{\dagger}}\bar{\partial}^{\tilde{\dagger}}(X_{(2)}+A_{(2)}) & =0,\\
\partial\bar{\partial}^{\tilde{\dagger}}(X_{(2)}+A_{(2)}) & =0,\\
\bar{\partial}\partial^{\tilde{\dagger}}(X_{(2)}+A_{(2)}) & =\ii\,\bar{\partial}\bar{\partial}^{\tilde{\dagger}}\imath_{\mu_{(2)}}\tilde{\omega}+\bar{\partial}\bar{\partial}^{\tilde{\dagger}}B_{(2)}.
\end{align}
As $X_{(2)}$ is a function of $x_{(2)}$ alone, we can perform gauge transformations of $x_{(2)}$ without affecting $A_{(2)}$ and $B_{(2)}$. Generically these gauge transformations will break the $\SU3$ normalisation condition, but we can always shift $\phi_{(2)}$ to compensate for this (which is what we have implicitly done by eliminating $\phi_{(2)}$ from the equations). An analogous argument to the one we gave previously then shows that we can always solve these conditions using the gauge freedom of $x_{(2)}$.

From this it is simple to see that this process can be continued to all orders. The conformally balanced condition at order $n$ is a set of equations for $x_{(n)}$ with $x_{(i<n)}$ fixed. Again, one can always find a gauge transformation of $x_{(n)}$ that solves these conditions. From this we conclude that the $D$-term conditions are gauge fixing conditions for the moduli and do not further constrain the moduli problem.

\subsection{Including bundle moduli}

We now want to show that when we include the bundle moduli we can solve the $D$-terms and Yang--Mills equations simultaneously. As stated at the start of section \ref{sec:par_defs}, we have the freedom to work with either $\End V$ or $\End V_{\mathbb{C}}$. We choose to work with $\End V_{\mathbb{C}}$ in this appendix and appendix \ref{app:massless}, so that the Donaldson--Uhlenbeck--Yau and Li--Yau theorems directly apply \cite{Donaldson85, UY86, YL87}.

The Yang–Mills condition is
\begin{equation}
\omega\wedge\omega\wedge F\propto\tilde{\rho}\wedge F=0.
\end{equation}
where we have defined $\tilde{\rho}=\tfrac{1}{2}\tilde{\omega}\wedge\tilde{\omega}$. A massless holomorphic deformation of this condition reads
\begin{equation}
\Delta_{0}\tilde{\rho}\wedge F+(\tilde{\rho}+\Delta_{0}\tilde{\rho})\wedge\partial_{A}\alpha_{0}=0.\label{eq:YM_gauge}
\end{equation}
Note that $\Delta_{0}\tilde{\rho}$ is completely determined by $x_{0}$. 

For now we assume $\text{H}^{0}(\End V_{\mathbb{C}})=0$. This is true for irreducible bundles with hermitian connections satisfying the Yang–Mills equation – such bundles are \emph{stable}. If $\text{H}^{0}(\End V_{\mathbb{C}})\neq0$ the Yang–Mills condition can give rise to Fayet--Iliopoulos $D$-terms in the lower-dimensional supergravity. Such terms have been studied extensively in heterotic compactifications (see \cite{AGL+09,Anderson:2011cza} and references therein). Their appearance can be understood in the context of the infinitesimal moduli problem as modding out by $\bar{D}$-exact terms~\cite{OS14b,OHS16}, see also appendix \ref{app:massless}. 

The equations of motion $\bar{D}y_{0}=0$ admit gauge transformations of $x_{0}$ and $\alpha_{0}$:
\begin{equation}
\begin{split}\delta x_{0} & =\partial a_{0}+\bar{\partial}b_{0}+\tr F\gamma_{0},\\
\delta\alpha_{0} & =\bar{\partial}_{A}\gamma_{0}.
\end{split}
\label{eq:bundle_gauge}
\end{equation}
Upon expanding $x_{0}$ and $\alpha_{0}$ in powers of $\epsilon$, the first order contributions to (\ref{eq:YM_gauge}) are
\begin{equation}
0=-\ii\,\ee^{-\phi}x_{(1)}\wedge\tilde{\omega}\wedge F+\tilde{\rho}\wedge\partial_{A}\alpha_{(1)}.
\end{equation}
We now want to show that we can solve this and the conformally balanced condition by an appropriate gauge choice. We denote these conditions schematically as
\begin{equation}
\bar{D}y_{(1)}=0,\qquad Cy_{(1)}=0,\qquad Yy_{(1)}=0,
\end{equation}
where $\bar{D}$ is the usual operator, and $C$ and $Y$ are the operators that act on $y_{(1)}$ to give the relevant equations. Explicitly (to first order) we have
\begin{align}
Cy_{(1)} & =\Bigl[\bar{\partial}^{\tilde{\dagger}}\bigl(\ee^{-\phi}(x_{(1)}-\tfrac{1}{2}\tilde{\omega}\lrcorner x_{(1)}\,\tilde{\omega})\bigr)\Bigr]+\Bigl[\partial^{\tilde{\dagger}}\bigl(\ee^{-\phi}(x_{(1)}-\tfrac{1}{2}\tilde{\omega}\lrcorner x_{(1)}\,\tilde{\omega})\bigr)-\ii\,\bar{\partial}^{\tilde{\dagger}}\imath_{\mu_{(1)}}\tilde{\omega}\Bigr],\\
Yy_{(1)} & =-\ii\,\ee^{-\phi}x_{(1)}\wedge\tilde{\omega}\wedge F+\tilde{\rho}\wedge\partial_{A}\alpha_{(1)}.
\end{align}

Let us start with a solution to the equations of motion
\begin{equation}
\bar{D}y_{(1)}=0.
\end{equation}
The equations of motion have a gauge symmetry under shifts by $\partial a+\bar{\partial}b+\tr F\gamma$:
\begin{equation}
\bar{D}y_{1}=0=\bar{D}(y_{(1)}+\partial a_{(1)}+\bar{\partial}b_{(1)}+\tr F\gamma_{(1)}).
\end{equation}
Generically a solution $y_{(1)}$ to $\bar{D}y_{(1)}=0$ will not solve either $Cy_{(1)}=0$ or $Yy_{(1)}=0$ on the nose. Let us make a gauge transformation of $y_{(1)}$:
\begin{equation}
y'_{(1)}=y_{(1)}+\partial\alpha_{(1)}+\bar{\partial}\beta_{(1)}.
\end{equation}
This still solves the equations of motion – $\bar{D}y'_{(1)}=0$ – but affects the remaining two conditions:
\begin{equation}
Cy'_{(1)}=Cy_{(1)}+C(\partial\alpha_{(1)}+\bar{\partial}\beta_{(1)}).
\end{equation}
We showed in the previous section that there is always a choice of $\alpha$ and $\beta$ that solves this condition, that is
\begin{equation}
Cy_{(1)}=-C(\partial\alpha+\bar{\partial}\beta)\quad\Rightarrow\quad Cy'_{(1)}=0.
\end{equation}
At this point we have a solution $y'_{(1)}$ that solves $\bar{D}y'_{(1)}=Cy'_{(1)}=0$ but generically has $Yy'_{(1)}\neq0$. Note also that the forms $\alpha$ and $\beta$ are fixed by $y_{(1)}$. We now want to show that we can always find a solution to $Yy=0$ using gauge transformations. Let us define
\begin{equation}
y''_{(1)}=y'_{(1)}+\partial a_{(1)}+\bar{\partial}b_{(1)}+\tr F\gamma_{(1)}+\bar{\partial}_{A}\gamma_{(1)}.
\end{equation}
This solves $\bar{D}y''_{(1)}=0$ – what happens to the other two conditions?
\begin{align}
Cy''_{(1)} & =C(\partial a_{(1)}+\bar{\partial}b_{(1)})+C(\tr F\gamma_{(1)}),\\
Yy''_{(1)} & =Yy'_{(1)}+Y(\partial a_{(1)}+\bar{\partial}b_{(1)})+Y(\tr F\gamma_{(1)})+Y(\bar{\partial}_{A}\gamma_{(1)}).
\end{align}
Again, we can solve $Cy''_{(1)}=0$ for \emph{any} $\gamma_{(1)}$ by choosing $a$ and $b$ appropriately. We now want to see if we can solve $Yy''_{(1)}=0$ using the gauge freedom in $\gamma_{(1)}$.

Note that $C$ picks out the $(\partial+\bar{\partial})$-exact terms of the input (and has no kernel acting on this) so that $Cy''_{(1)}=0$ is equivalent to
\begin{equation}
\partial a_{(1)}+\bar{\partial}b_{(1)}+\bigl[\tr F\gamma_{(1)}\bigr]_{\partial+\bar{\partial}}=0.
\end{equation}
Using this we have
\begin{equation}
Yy''_{(1)}=Yy'_{(1)}+Y(\mathcal{P}[\tr F\gamma_{(1)}])+Y(\bar{\partial}_{A}\gamma_{(1)}),
\end{equation}
where $\mathcal{P}$ projects onto $\Omega^{(1,1)}(X)\backslash\{\image\partial+\image\bar{\partial}\}$. Explicitly we have\footnote{Here we have used $\tilde{\star}1=\tfrac{1}{6}\tilde{\omega}^{3}$, $\tilde{\star}(a_{11}\wedge\tilde{\omega}\wedge b_{11})=-b_{11}^{\tilde{\sharp}}\lrcorner a_{11}$ if $b_{11}$ is a primitive $(1,1)$-form (like $F$), and $\tilde{\star}(\tilde{\rho}\wedge a_{11})=\tilde{\omega}\lrcorner a_{11}$. We use $\tilde{\sharp}$ to denote raising with the Gauduchon metric.}
\begin{equation}
\label{eq:YMdef1}
Yy''_{(1)}=\ii\,\tilde{\star}\,\bigl(\ee^{-\phi}F^{\tilde{\sharp}}\lrcorner x'_{(1)}+\bar{\partial}_{A}^{\tilde{\dagger}}\alpha'_{(1)}+\ee^{-\phi}F^{\tilde{\sharp}}\lrcorner\mathcal{P}[\tr F\gamma_{(1)}]+\tilde{\Delta}_{A}\gamma_{(1)}\bigr).
\end{equation}
One can check that the operator acting on $\gamma_{(1)}$ is a positive semi-definite elliptic operator with trivial kernel, so one can always find a $\gamma_{(1)}$ that solves $Yy''_{(1)}=0$. Again, one can see that this argument will hold to all orders in $\epsilon$.

Recall that we included $TX$ as part of the gauge bundle $V$. Strictly speaking one should show you can solve the analogue of \eqref{eq:YMdef1} for $TX$ as well. Naively an extra minus sign will appear in the trace term, leading to a question of whether the operator acting on $\gamma_{(1)}$ has trivial kernel. Physically we expect this equation will not cause any problems -- there is evidence that the degrees of freedom in the connection for $TX$ are not real moduli as they can be eliminated using field redefinitions (at least to $\mathcal{O}(\alpha')$)~\cite{OS14}. Another way of saying this is that the hermitian Yang--Mills condition for $TX$ is automatically satisfied provided the other supersymmetry conditions are solved and one chooses the appropriate connection, which at $\mathcal{O}(\alpha')$ is the connection $\nabla^-$. Though we do not have a proof, we expect this will hold to higher order in $\alpha'$ when one chooses the connection on $TX$ to be an instanton.

\subsection{Polystable bundles}

Let us now consider the situation where the bundle is {\it polystable}, that is we have a reducible bundle where each factor is stable of slope zero. If a bundle factor has non-vanishing first Chern class, which is possible for a sum of irreducible components, then it admits a single section proportional to the identity isomorphism on the bundle factor. This can give rise to a non-trivial Fayet--Iliopoulos term in the lower-dimensional supergravity. We now show that equation \eqref{eq:YMdef1} can still be solved when such sections are present. 

A polystable bundle is a sum of irreducible components
\begin{equation}
V_{\mathbb{C}}=\bigoplus_iV^{(i)}_{\mathbb{C}},
\end{equation}
where each factor $V^{(i)}_{\mathbb{C}}$ is a stable slope-zero bundle with non-zero trace. The trace-less part of \eqref{eq:YMdef1} can be solved by the methods outlined in the previous subsection. The non-trivial part comes from the trace, which we isolate by taking the trace to get
\begin{equation}
\label{eq:YMdef2}
-\ii\,\tilde{\star}\tr(Yy''_{(1)})^{(i)}=\ee^{-\phi}\tr F^{(i)\,\tilde{\sharp}}\lrcorner x'_{(1)}+\bar{\partial}^{\tilde{\dagger}}_A \tr\alpha'_{(1)}+\ee^{-\phi}\tr F^{(i)\,\tilde{\sharp}}\lrcorner\mathcal{P}[\tr F^{(i)}\gamma^{(i)}_{(1)}]+\tilde{\Delta}\tr\gamma^{(i)}_{(1)}.
\end{equation}
It is simplest to show that this can be solved by reintroducing the $\alpha'$ parameter, which we have neglected until this point. To simplify the notation, we imagine the bundle $V_{\mathbb{C}}$ is abelian for the remainder of the section. Reintroducing $\a$, equation \eqref{eq:YMdef2} becomes
\begin{equation}
\label{eq:YMdef3}
-\ii\,\tilde{\star}\,Y y''_{(1)}=\ee^{-\phi}F^{\tilde{\sharp}}\lrcorner x'_{(1)}+\bar{\partial}_A^{\tilde{\dagger}}\alpha'_{(1)}+\tfrac{1}{4}\a\,\ee^{-\phi}\,F^{\tilde{\sharp}}\lrcorner\mathcal{P}[F\gamma_{(1)}]+\tilde{\Delta}\gamma_{(1)}.
\end{equation}
We can now expand this equation in $\a$. We expand the deformations $x'_{(1)}$ and $\alpha'_{(1)}$, together with the gauge  parameter $\gamma_{(1)}$, while keeping the background geometry fixed. The potential non-zero contributions to $\gamma_{(1)}$ are then
\begin{equation}
\gamma_{(1)}=\frac{1}{\a}\,\gamma^{(-1)}_{(1)}+\gamma_{(1)}^{(0)}+\a\,\gamma_{(1)}^{(1)}+\ldots
\end{equation}
It follows from \eqref{eq:YMdef3} that $\gamma^{(-1)}_{(1)}$ must be constant. The expression we get at zeroth order in $\a$ is
\begin{equation}
\label{eq:YMdef4}
-\ii\,\tilde{\star}\,Y{y''}^{(0)}_{(1)}=\ee^{-\phi}F^{\tilde{\sharp}}\lrcorner {{x'}^{(0)}_{(1)}}+\bar{\partial}^{\tilde{\dagger}}_A{\alpha'}^{(0)}_{(1)}+\tfrac{1}{4}\,\ee^{-\phi}\,F^{\tilde{\sharp}}\lrcorner F\,\gamma^{(-1)}_{(1)}+\tilde{\Delta}_A\gamma_{(1)}^{(0)}=0,
\end{equation}
where we note that 
\begin{equation}
\mathcal{P}[F\gamma^{(-1)}_{(1)}]=F\gamma^{(-1)}_{(1)},
\end{equation}
for constant $\gamma^{(-1)}_{(1)}$. The constant part of \eqref{eq:YMdef4} can now be obtained by integrating over $X$. Doing so, we get
\begin{equation}
-\ii\int_X\tilde{\star}\,Y{y''}^{(0)}_{(1)}=\int_X\ee^{-\phi}F^{\tilde{\sharp}}\lrcorner {{x'}^{(0)}_{(1)}}+\tfrac{1}{4}\gamma^{(-1)}_{(1)}\int_X \ee^{-\phi}\,F^{\tilde{\sharp}}\lrcorner F,
\end{equation}
This determines $\gamma^{(-1)}_{(1)}$ as a function of ${{x'}^{(0)}_{(1)}}$. The remaining non-constant part can then be solved for modulo a constant term, which determines $\gamma_{(1)}^{(0)}$.

It is easy to see that this can be continued to higher orders in $\a$. At the next order, we use the so-far undetermined constant piece of $\gamma_{(1)}^{(0)}$ to fix the constant piece of the first-order equation. This determines $\gamma_{(1)}^{(0)}$, which is now independent of $\gamma_{(1)}^{(1)}$. We then solve for the non-constant piece, determining $\gamma_{(1)}^{(1)}$ modulo a new constant. It is also easy to check that a similar argument will hold at higher orders in the expansion parameter $\epsilon$.

The argument we have given here for a single abelian factor is easily generalised to the general polystable case. This shows that the $D$-term equations can be solved even when there are Fayet--Iliopoulos terms present. The reader may have expected that the $D$-term equations would obstruct some hermitian moduli when FI terms are present. This is not quite right; instead the expected obstructed moduli will form part of the gauge transformations and so should not be included in the first place. We give more detail on this in appendix \ref{app:massless}.

\subsection{Full Maurer--Cartan equations}

The arguments we have given for the massless deformations can be generalised to solutions to the full set of Maurer–Cartan equations:
\begin{align}
\bar{D}y-\tfrac{1}{2}[y,y]-\tfrac{1}{2}\partial b & =0,\\
\bar{\partial}b-\tfrac{1}{2}y^{a}\partial_{a}b+\tfrac{1}{3!}\langle y,[y,y]\rangle & =0,\\
\partial\imath_{\mu}\Omega & =0.
\end{align}
What is the gauge symmetry of these equations? The $L_{3}$ algebra structure gives us the gauge transformation of $Y$ by a gauge parameter $\Lambda=(\lambda,\xi)\in\mathcal{Y}_{0}$
\begin{equation}
\begin{split}
\delta_{\Lambda}Y & =\ell_{1}(\Lambda)+\ell_{2}(\Lambda,Y)-\tfrac{1}{2}\ell_{3}(\Lambda,Y,Y)\\
 & =\bigl(\bar{D}\lambda+\tfrac{1}{2}\partial\xi+[\lambda,y],0\bigr)\\
 & \eqspace+\bigl(0,\bar{\partial}\xi+\tfrac{1}{2}(\langle\lambda,\partial b\rangle-\langle y,\partial\xi\rangle)-\tfrac{1}{3}(\tfrac{1}{2}\langle\lambda,[y,y]\rangle+\langle y,[y,\lambda]\rangle)\bigr).
\end{split}
\end{equation}
Now expand $Y$ and $\Lambda$ in powers of $\epsilon$. At first order the equations of motion and gauge transformations are
\begin{align}
\bar{D}y_{(1)}-\tfrac{1}{2}\partial b_{(1)} & =0, & \delta y_{(1)} & =\bar{D}\lambda_{(1)}+\tfrac{1}{2}\partial\xi_{(1)},\\
\bar{\partial}b_{(1)} & =0, & \delta b_{(1)} & =\bar{\partial}\xi_{(1)},\\
\partial\imath_{\mu_{(1)}}\Omega & =0.
\end{align}
Exactly as before, $x_{(1)}$ and $\alpha_{(1)}$ have a gauge symmetries that allow us to solve the $D$-term conditions to first order.

At second order we have
\begin{align}
\bar{D}y_{(2)}-\tfrac{1}{2}[y_{(1)},y_{(1)}]-\tfrac{1}{2}\partial b_{(2)} & =0, & \delta y_{(2)} & =\bar{D}\lambda_{(2)}+\tfrac{1}{2}\partial\xi_{(2)}+[\lambda_{1},y_{1}],\\
\bar{\partial} b_{(2)}-\tfrac{1}{2}y_{(1)}^{a}\partial_{a}b_{(1)} & =0, & \delta b_{(2)} & =\bar{\partial}\xi_{(2)}+\tfrac{1}{2}\bigl(\langle\lambda_{(1)},\partial b_{(1)}\rangle-\langle y_{(1)},\partial\xi_{(1)}\rangle\bigr),\\
\partial\imath_{\mu_{(2)}}\Omega & =0.
\end{align}
Here $(y_{(1)}, b_{(1)})$ are the gauge-transformed fields that solve the first-order $D$-term conditions. Furthermore $\lambda_{1}$ is fixed by these first-order conditions. We note that the second order field $y_{(2)}$ is shifted by the first order gauge transformation by the term $[\lambda_{1},y_{1}]$. This means it is $y_{(2)}-[\lambda_{(1)},y_{(1)}]$ that appears in the second-order $D$-term condition, which then admits shifts by terms of the form $\bar{D}\lambda_{(2)}+\tfrac{1}{2}\partial\xi_{(2)}$ which we may use to solve the $D$-term conditions just as in the first order case, fixing $\lambda_{(2)}$ and $\xi_{(2)}$ in the process. We can see that this structure will continue to hold to all orders. Assuming we have solved the $D$-terms at $i^{\text{th}}$ order where $i<n$, $y_{(n)}$ is shifted by gauge transformations involving the $y_{(i)}$. We can then use the shifts of $y_{(n)}$ by $(\partial+\bar{\partial})$-exact forms to solve the $D$-terms at $n^{\text{th}}$ order. Solving the $D$-term conditions thus fixes the gauge symmetries (though there are residual gauge-of-gauge transformations).

%%%%%%%%%%%%%%%%%%%%%%%%%%%%%%%%%%%%%%%%%%%%%%%%%%%%%%%%%%%%%%%%%%%%%%%%%%

\section{Massless moduli}
\label{app:massless}

In this appendix we clarify the meaning of the massless moduli
\begin{equation}
[y_0]\in \HH^{(0,1)}_{\bar D}(\tilde \Q),\quad y_0=(\mu_0,\alpha_0,x_0).
\end{equation}
The massless moduli have been analysed in great detail in the \cite{AGS14, OS14b, GRT15} -- what follows is a short review. The first thing to notice is that 
\begin{equation}
\bp\mu_0=0.
\end{equation}
The fields $\mu_0$ parametrise the complex structure moduli which, modulo gauge symmetries, take values in the cohomology
\begin{equation}
[\mu_0]\in \HH^{(0,1)}_{\bp}(T^{(1,0)}X)\cong \HH^{(2,1)}_{\bp}(X).
\end{equation}
Recall that we also impose the extra condition
\begin{equation}
\p\imath_{\mu_0}\Omega=0.
\end{equation}
This condition can be viewed as an equation of motion for the field $b$ appearing in the superpotential \eqref{eq:action_with_b}. It is also a necessary condition for preserving the $\d$-closure of the holomorphic three-form, which in general is a stronger condition then preserving an integrable complex structure alone. However, with the extra assumption that $\HH^{(0,1)}_{\bp}(X)=0$, it can be shown that each class in $ \HH^{(0,1)}_{\bp}(T^{(1,0)}X)$ has a representative satisfying this condition \cite{OS14b}.

Next we must solve the equation 
\begin{equation}
\label{eq:AtiyahFull}
\bar{\partial}_{A}\alpha_0+F_{b}\wedge\mu_0^{b}=0 .
\end{equation}
This is then the familiar condition that the complex structure deformations $\mu^a$ should be in the kernel of the Atiyah map
\begin{equation}
{\cal F}_\mu\colon \HH^{(0,1)}_{\bp}(T^{(1,0)}X)\rightarrow \HH^{(0,2)}_{\bp_A}(\End V_{\mathbb{C}}),
\end{equation}
where ${\cal F}_\mu$ is given by contraction with $F$ as in \eqref{eq:AtiyahFull}. The fields $\alpha_0$ now decompose as
\begin{equation}
\alpha_0=\alpha_0^\text{c}+\alpha_0^{\mu_0},
\end{equation}
where $\alpha_0^\text{c}$ is closed and $\alpha_0^{\mu_0}$ is uniquely determined by the complex structure moduli through equation \eqref{eq:AtiyahFull}. Modulo gauge symmetries, we have
\begin{equation}
[\alpha_0^\text{c}]\in \HH^{(0,1)}_{\bp_A}(\End V_{\mathbb{C}}),
\end{equation}
which we refer to as gauge moduli.  

The final equation we must solve is
\begin{equation}
\label{eq:Atiyah2Full}
\bar{\partial}x_{0\,a}+\ii(\partial\omega)_{ea\bar{c}}\,\d z^{\bar{c}}\wedge\mu_0^{e}-\tr(F_{a}\wedge\alpha_0)=0.
\end{equation}
Note that we need only solve this equation modulo $\p_a$-exact two-forms. In particular, the equation we should solve is 
\begin{equation}
\label{eq:Atiyah2Full2}
\bar{\partial}x_{0\,a}+\ii(\partial\omega)_{ea\bar{c}}\,\d z^{\bar{c}}\wedge\mu_0^{e}-\tr(F_{a}\wedge\alpha_0)=\p_ab ,
\end{equation}
for some $b\in\Omega^{(0,2)}(X)$ which depends on the particular sheaf representative. Note however that if $\HH_{\p}^{(0,2)}(X)\neq0$, then we can always add an element of $\HH_{\p}^{(0,2)}(X)$ to $b$ without changing this equation. Hence $\HH_{\p}^{(0,2)}(X)$, i.e.~anti-holomorphic sections of $\Omega^{(0,2)}(X)$, should in principle be considered as part of the massless moduli. We have throughout this paper assumed that this cohomology vanishes. 

Let us define the maps
\begin{align}
{\cal H}&\colon\ker{\cal F}\rightarrow \HH^{(0,2)}(T^{*(1,0)}X\,/\,{\p\textrm{-exact}}),\\
{\cal F}_\alpha&\colon \HH^{(0,1)}_{\bp_A}(\End V_{\mathbb{C}})\rightarrow \HH^{(0,2)}(T^{*(1,0)}X\,/\,{\p\textrm{-exact}}).
\end{align}
The maps are given by
\begin{align}
{\cal H}(\mu_0)_a&=\ii(\partial\omega)_{ea\bar{c}}\,\d z^{\bar{c}}\wedge\mu_0^{e}-\tr(F_{a}\wedge\alpha^{\mu_0}_0),\\
{\cal F}_\alpha(\alpha^\text{c}_0)_a&=-\tr(F_{a}\wedge\alpha^\text{c}_0).
\label{eq:KerBundle}
\end{align}
It is easy to check that the right-hand side of both of these equations is $\bp$-closed. The condition imposed by \eqref{eq:Atiyah2Full} is that the sum should be exact. The number of remaining massless moduli is then given by
\begin{equation}
\vert\ker{\cal H}\vert+\vert\ker{\cal F}_\alpha\vert+\vert\image{\cal H}\cap\image{\cal F}_\alpha\vert.
\end{equation}
We have the freedom to view the extra moduli in the last term as either complex structure moduli, bundle moduli, or something in between. Viewing it as complex structure moduli means that the moduli $\mu_0$ which remain massless are those which satisfy
\begin{equation}
\label{eq:Hmassless}
{\cal H}(\mu_0)\subseteq\image{\cal F}_\alpha .
\end{equation}
The remaining massless bundle moduli $\alpha_0^\text{c}$ then need be in the kernel of ${\cal F}_\alpha$. 

Equation \eqref{eq:Atiyah2Full} now determines the non-closed part of $x_0$ uniquely in terms of the massless complex structure and bundle moduli. 
The remaining hermitian moduli $x_0^\text{c}$ are $\bp$-closed modulo $\p_a$-exact terms, and hence take values in the cohomology
\begin{equation}
[x_0^\text{c}]\in \HH^{(0,1)}_{\bp}(T^{*(1,0)}X\,/\,\p\textrm{-exact}).
\end{equation}
These are the fields conventionally referred to as K\"ahler moduli or hermitian moduli. As it turns out, this cohomology forms a subset of the Aeppli cohomology \cite{Aeppli65}.
Indeed, in order to define an element of $\HH^{(0,1)}_{\bp}(T^{*(1,0)}X\,/\,\p\textrm{-exact})$, we need that
\begin{equation}
\label{eq:x0Coh}
\bp x_0=\p\kappa,
\end{equation}
for some $\kappa\in\Omega^{(0,2)}(X)$. In particular, $\p\bp x_0=0$. Modulo the gauge symmetries of $x_0$ -- shifts by $\bp$- and $\p$-exact terms -- we see that $x_0$ is an element in the Aeppli cohomology.

For the geometries we are considering in this appendix, it is true that $\HH^{(0,1)}_{\bp}(T^{*(1,0)}X)$ is isomorphic to $\HH^{(1,1)}_{\bp}(X)$. Let us see why. Note that \eqref{eq:x0Coh} implies that $\bp\kappa=0$, which by $\HH^{(0,2)}(X)=0$ implies that $\kappa=\bp\beta$ for some $\beta\in\Omega^{(0,1)}(X)$. Let us now use the freedom to shift $x_0$ by a $\p$-exact term $\p\alpha$ to get
\begin{equation}
\bp x_0=\p(\bp\kappa+\bp\alpha)\:.
\end{equation}
We can choose $\alpha$ such that the right-hand side vanishes. This is equivalent to $\bp\kappa+\bp\alpha=0$, which by the Hodge decomposition \eqref{eq:exactHodge} fixes the $\partial\bar{\partial}^{\tilde{\dagger}}$-exact part of $\p\alpha$. The remaining $\p$-exact part of $\p\alpha$ is also $\bp$-exact. We conclude that for the geometries under consideration, we have
\begin{equation}
\HH^{(0,1)}_{\bp}(T^{*(1,0)}X\,/\,\p\textrm{-exact})\cong \HH^{(0,1)}_{\bp}(T^{*(1,0)}X)=\HH^{(1,1)}_{\bp}(X)\:,
\end{equation}
and the hermitian moduli are counted by $h^{(1,1)}$ as in a Calabi--Yau compactification. 

We now compare this with the cohomology $\HH^{(0,1)}_{\bar D}(\tilde{\cal Q})$. Using long exact sequences in cohomology, it is a straightforward exercise to show that
\begin{equation}
\HH^{(0,1)}_{\bar D}(\tilde{\cal Q})\cong\frac{\HH^{(0,1)}_{\bp}(T^{*(1,0)}X\,/\,\p\textrm{-exact})}{\image{\cal F}_\alpha}\oplus\ker({\cal H}+{\cal F}_\alpha).
\end{equation}
This is almost as expected, except that we are modding out the hermitian moduli by elements in $\image{\cal F}_\alpha$, where we now view ${\cal F}_\alpha$ as
\begin{equation}
{\cal F}_\alpha\colon\HH^0(\End V_{\mathbb{C}})\rightarrow\HH^{(0,1)}_{\bp}(T^{*(1,0)}X\,/\,\p\textrm{-exact}).
\end{equation}
We can understand this from considering gauge transformations by a $\bar D$-exact term. For polystable bundles, part of these transformations are formed by sections of $\HH^0(\End V_{\mathbb{C}})$. Again we have $V_{\mathbb{C}}=\bigoplus_i V^{(i)}_{\mathbb{C}}$ for stable slope-zero factors $V^{(i)}_{\mathbb{C}}$. Thus we have 
\begin{equation}
\image{\cal F}_\alpha=\text{span}\{c_1 F_1+\ldots+c_n F_n\},
\end{equation}
for constants $c_i$ such that $\sum_in_ic_i=0$, where $n_i$ is the dimension of the $i^{\text{th}}$ bundle, so that the total section remains traceless. As we saw in the previous section, these are precisely the gauge transformations needed to solve the $D$-term conditions in the presence of Fayet--Iliopoulos terms, and so hermitian elements of this form should not be thought of as moduli.

\end{document}